\documentclass[a4paper,fleqn]{cas-sc}

\usepackage[authoryear,longnamesfirst]{natbib}
\usepackage{graphicx}
\usepackage{epstopdf, epsfig}
\usepackage{physics,cancel,mathtools}
\usepackage{bm,xcolor}
\usepackage{caption,subcaption,calc}

\def\tsc#1{\csdef{#1}{\textsc{\lowercase{#1}}\xspace}}
\tsc{WGM}
\tsc{QE}

\newtheorem{recipe}{Numerical recipe}
\newcommand{\iu}{\mathrm{i}}
\newcommand{\dir}[1]{\hat{\bm{#1}}}
\newcommand{\mathsfbi}[1]{\bm{\mathsf{#1}}}

\begin{document}
\let\WriteBookmarks\relax
\def\floatpagepagefraction{1}
\def\textpagefraction{.001}

\shorttitle{}    

\shortauthors{}  

\title [mode = title]{A semi-analytical pseudo-spectral method for 3D Boussinesq equations of rotating, stratified flows in unbounded cylindrical domains}



%
\author[1]{Jinge Wang}[linkedin={jinge-wang-cfd},
orcid=0009-0008-5310-8230]
\ead{jinge@berkeley.edu}
\credit{Conceptualization, Methodology, Software, Validation, Formal analysis, Investigation, Writing - Original Draft, Writing - Review \& Editing, Visualization}

\author[1]{Philip S. Marcus}[orcid=0000-0001-5247-0643]
\cormark[1]
\ead{pmarcus@me.berkeley.edu}
\ead[url]{https://cfd.me.berkeley.edu}
\credit{Methodology, Writing - Review \& Editing, Supervision, Funding acquisition}

\cortext[1]{Corresponding author}



\begin{abstract}
We present a pseudo-spectral method for solving the three-dimensional Boussinesq equations in unbounded cylindrical domains, specifically tailored for rotating, stably stratified flows subject to strong azimuthal shear. To effectively capture the global geometry without sacrificing spectral accuracy, the spatial discretization employs Fourier expansions in the azimuthal and axial directions alongside mapped associated Legendre polynomials in the radial direction. This basis spans the semi-infinite domain while analytically resolving the coordinate singularity at the origin. While this spectral framework ensures high spatial fidelity, the temporal integration of these rotating shear flows presents a formidable computational challenge due to the numerical stiffness driven by fast restorative wave forces and rapid background advection. To circumvent this, we develop an exponential time differencing (ETD) scheme that analytically integrates the fully coupled linear operator, including the radially dependent advective cross terms. By encoding the physical resonance characteristics and stability limits of the background flow directly into the integration operators, the proposed ETD formulation removes the numerical stability constraints imposed by the background shear and stratification. This permits integration time steps scaled by the slow macroscopic evolution of the physical instabilities rather than the fast background kinematics, offering significant performance gains over standard mixed implicit-explicit schemes. The method's accuracy and stability are validated through the precise conservation of energy and angular momentum, establishing a robust framework for simulating instabilities in astrophysical and geophysical vortices.
\end{abstract}



\maketitle

\section{Introduction}

\begin{figure}
    \centering
    \includegraphics[height=0.22\textheight]{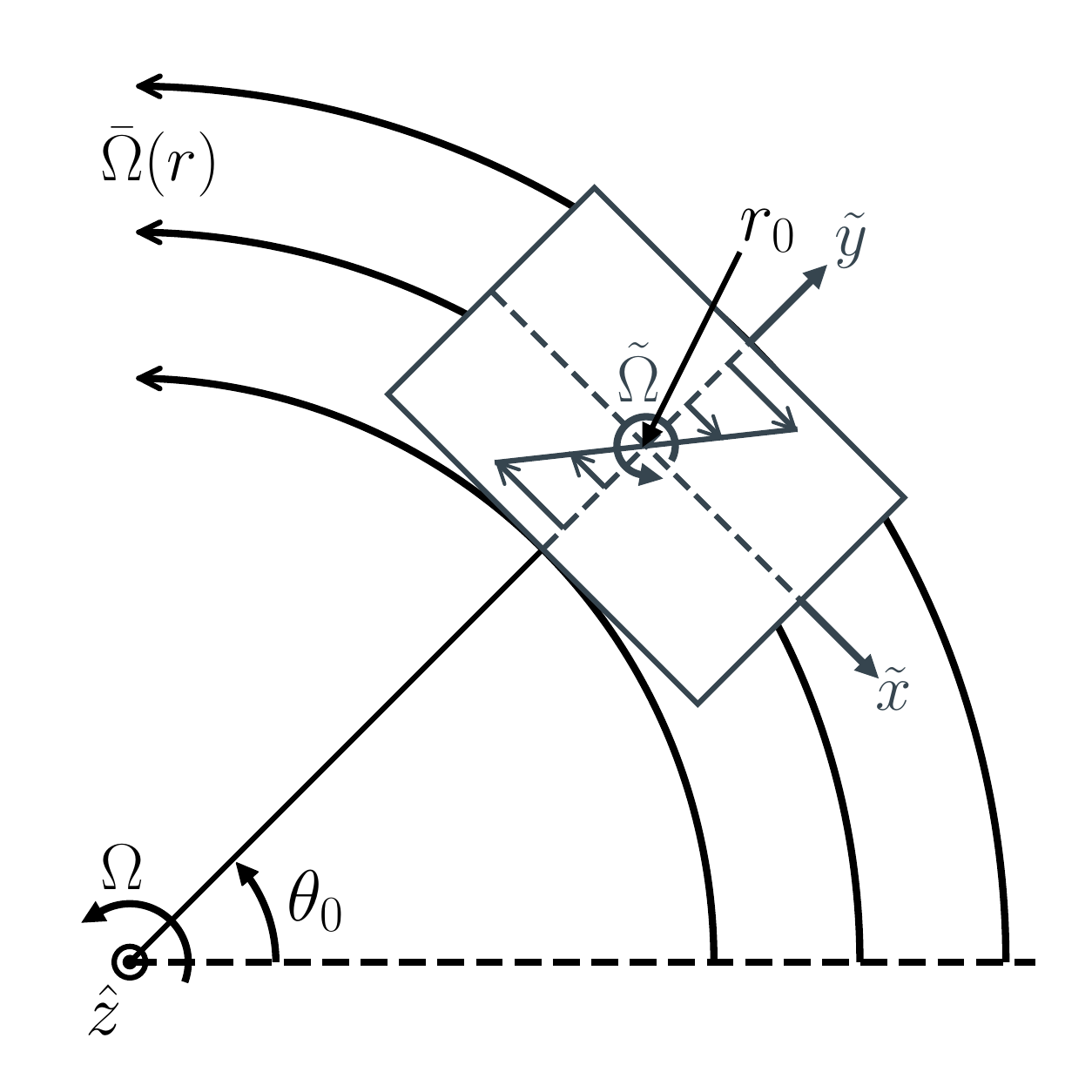}
    \caption{Mapping between the global cylindrical coordinates $(r, \theta, z)$ and the local shearing box $(\tilde{x}, \tilde{y}, \tilde{z})$ centered at $(r_0, \theta_0)$. The local coordinates are defined as $\tilde{y} = r - r_0$, $\tilde{x} = -r_0(\theta - \theta_0)$, and $\tilde{z} = z$, such that the local unit vectors correspond to $\tilde{\bm{y}} = \dir{r}$ and $\tilde{\bm{x}} = -\dir{\theta}$. Gravity acts in the $-z$ direction (pointing into the page). The local frame co-rotates with the background at an angular velocity $\tilde{\Omega} = \bar{\Omega}(r_0) + \Omega$, where $\bar{\Omega}$ represents the radially dependent angular velocity of the global background flow (typically following a power-law profile, such as the Keplerian $\bar{\Omega}\propto r^{-3/2}$ in accretion disks) and $\Omega$ is the constant background rotation rate of the global frame. The differential rotation of the background flow is locally approximated as a linear shear profile $\bar{u}_{\tilde{x}} = -\sigma \tilde{y}$, where the constant shear rate is defined as $\sigma = r_0 (\dd\bar{\Omega}/\dd{r})|_{r_0}$ ($\sigma < 0$ in the depicted example).}
    \label{fig:shearing_box_schematic}
\end{figure}

In geophysical and astrophysical systems, ranging from planetary atmospheres to protoplanetary accretion disks, the transport of momentum and the dissipation of energy are often governed by the coexistence of strong background rotation, stable density stratification, and differential azimuthal shear \citep{Barranco_2005, Armitage_2011}. The combined action of these three physical effects can trigger complex hydrodynamic phenomena, including the stratorotational instability (SRI) \citep{Molemaker_2001,Yavneh_2001,Dubrulle_2005} and the zombie vortex instability (ZVI) \citep{Marcus_2013, Marcus_2015}. Historically, numerical investigations and analytical advances characterizing these systems have relied on the local shearing box approximation (see Fig. \ref{fig:shearing_box_schematic}) \citep{Balbus_1998, barranco_2006}. However, this Cartesian idealization possesses inherent mathematical and physical limitations \citep{regev_2008}. Specifically, the shearing box artificially decouples the local shear rate from the local background rotation, neglects global curvature terms of order $\mathcal{O}(1/r)$, and imposes a strictly linear background velocity profile that can artificially restrict the macroscopic radial evolution of large-scale modes. Consequently, while phenomena like the ZVI manifest within the shearing box, their viability and structural evolution in a continuous global domain remain open questions. To enforce geometric constraints and investigate whether such instabilities can be sustained physically, it is necessary to study the system in a global domain \citep{Armitage_1998, Machida_2000, Ebrahimi_2016, Zhu_2020}.

To discretize these global domains spatially, spectral methods are frequently employed. By projecting the spatial domain onto a basis of orthogonal continuous functions and evaluating spatial derivatives analytically in spectral space, spectral methods offer exponential convergence and minimal numerical dissipation for smooth functions \citep{canuto_2007}. Modern open-source frameworks, such as the Dedalus project \citep{Dedalus}, have demonstrated the flexibility and widespread application of these techniques for fluid modeling. To study the system in a global cylindrical domain, however, standard polynomial bases require specialized parity constraints to handle the coordinate singularity at the origin \citep{Boyd_2011}. Furthermore, for many astrophysical applications, it is beneficial to utilize an unbounded cylindrical domain to isolate the fluid from the artificial wave reflections associated with rigid outer boundaries. To achieve this, the present framework utilizes mapped associated Legendre polynomials for the semi-infinite radial domain \citep{matsushima_1997}, combined with Fourier expansions in the azimuthal and axial directions. Exhibiting natural analyticity at the coordinate singularity and algebraic decay at infinity, this combination forms a robust Galerkin basis that ensures high computational efficiency without requiring explicit enforcement of pole conditions.

To model the thermodynamics and kinematics of stratified systems, the Boussinesq approximation is widely utilized \citep{Spiegel_1960}. By filtering out acoustic waves, this approximation renders the velocity field strictly divergence-free. This constraint is typically enforced via a poloidal-toroidal decomposition, which also analytically eliminates the pressure potentials from the momentum equations \citep{matsushima_1997,lee_2023,wang_2024}. For time integration, the Boussinesq equations are routinely advanced using pseudo-spectral implicit-explicit (IMEX) schemes, which treat linear diffusive terms implicitly in spectral space (e.g., via Crank-Nicolson) and nonlinear advective and Coriolis terms explicitly in the physical collocation space (e.g., via Adams-Bashforth) \citep{Garcia_2010,Garcia_2014,Gardner_2018,Gopinath_2022}.  However, the combined physical effects of strong rotation, stable stratification, and intense azimuthal shear introduce significant numerical stiffness. The rapid advective transport driven by the non-hydrostatic background flow and the fast restorative inertial-gravity (Poincaré) waves impose restrictive Courant-Friedrichs-Lewy (CFL) conditions \citep{Gardner_2018,Vogl_2019}, which force the explicit time step to be impractically small to maintain numerical stability. This computational bottleneck becomes particularly prohibitive for slow instabilities, such as the ZVI, which evolve over long timescales \citep{Wohlever_2025} and require large numerical time steps to remain computationally tractable.

To circumvent the combined stiffness of the restorative wave forces and the background advection without relying on expensive global factorizations, exponential time differencing (ETD) methods offer an effective alternative \citep{Cox_2002,Kassam_2005,Garcia_2014}. ETD methods utilize matrix exponentials to analytically integrate the fully coupled linear portion of the governing equations, natively stepping over the fastest linear timescales. While historically limited by algorithmic complexity, as discussed by \citep{Moler_2003}, the development of stabilization techniques such as contour integration and Krylov subspace approximations has mitigated issues with exponentiating large-scale matrices \citep{Schulze_2009,Lan_2021}. In the present study, the specific choice of the mapped associated Legendre basis structures the linear operator into independent, low-dimensional blocks, which facilitates a straightforward ETD implementation. While the poloidal-toroidal decomposition is highly advantageous for standard IMEX schemes, it complicates the exact matrix diagonalization required by ETD. Consequently, the proposed ETD scheme is formulated using primitive velocity variables, and the poloidal-toroidal decomposition is utilized subsequently as an exact projection method to act as a pressure solver and enforce the divergence-free condition.

The remainder of this paper is organized as follows. In \S\ref{sec:governing-eqns}, we present the fully nonlinear governing equations for a rotating, stratified Boussinesq fluid and formulate the specific global background shear profiles. In \S\ref{sec:numerical-methods}, we detail the spatial discretization using the mapped associated Legendre basis and construct the basic block-diagonal ETD time-stepping framework. We further extend the ETD scheme to analytically diagonalize the background advective shear, detailing its physical connections to centrifugal stability and critical layer resonance. Finally, in \S\ref{sec:numerical-tests}, we present numerical validations of the code against known linear stability bounds, followed by conclusions in \S\ref{sec:conclusions}.

\section{Governing equations}\label{sec:governing-eqns}
\subsection{Boussinesq equations with azimuthal shear}
We consider the dynamics of a viscous, stratified fluid within an unbounded global cylindrical domain defined by coordinates $(r, \theta, z)$ with unit vectors $(\dir{r}, \dir{\theta}, \dir{z})$, where $0 \leq r < \infty$. The fluid is subject to a constant gravitational acceleration $\bm{g} = -g\dir{z}$, and the system is observed in a frame of reference rotating about the vertical axis with constant angular velocity: $\bm{\Omega} = \Omega\dir{z}$, as depicted in figure~\ref{fig:shearing_box_schematic}. The primitive state variables of the fluid are its velocity $\bm{u}$, gauge pressure $p$, and rescaled buoyancy $b \equiv \rho g / {\rho_0}$. The latter two are defined relative to a hydrostatic reference state of constant density $\rho_0$. 

To isolate the dynamics of interest, we decompose the total flow state into a steady, axisymmetric background component, denoted by overbars, and a time-dependent component, denoted by primes (e.g., $p = \bar{p} + p'$). 
The background state is considered to be an azimuthal shear flow with radial dependence, 
\begin{equation}
    \bar{\bm{U}} \equiv \bar{U}_\theta(r)\dir{\theta},
    \label{eqn:baseflow}
\end{equation}
whose angular velocity $\bar{\Omega}$ and axial vorticity $\bar{\zeta}$ can be expressed as
\begin{subequations}
\begin{equation}\label{eqn:baseflow-vorticity}
    \bar{\Omega}(r) = \frac{\bar{U}_\theta}{r}, \quad
    \bar{\zeta}(r) = \frac{1}{r}\dv{(r\bar{U}_\theta)}{r} = 2\bar{\Omega} + r\dv{\bar{\Omega}}{r}.
    \tag{\theequation a,b}
\end{equation}
\end{subequations}
The background buoyancy is assumed to depend solely on the vertical coordinate and follow a linear profile, characterized by a constant Brunt-Väisälä frequency 
\begin{equation}
    \bar{N} = \sqrt{-\dv{}{z}\bar{b}(z)}.
\end{equation} 
We further require the disturbance component to remain analytic at origin ($r=0$) and to decay sufficiently rapidly to zero in the far field ($r\rightarrow\infty$), so that it is physically confined and that both its total mass anomaly and energy are finite. A more detailed discussion on these pole conditions is presented in \S\ref{sec:PT-decomp}.


We adopt the Boussinesq approximation to govern the flow dynamics, assuming that density variations are sufficiently small relative to $\rho_0$ that they are negligible in the inertial terms but retained in the gravitational force. Under this approximation, the specified background state constitutes an exact equilibrium solution in the inviscid limit, as long as the background pressure field $\bar{p}(r,z)$ satisfies the hydrostatic and cyclogeostrophic balances:
\begin{subequations}
\begin{align}
    \partial_r\bar{p} &= \rho_0(\bar{\Omega}+2\Omega)\bar{U}_\theta, \\
    \partial_\theta\bar{p} &= 0, \\
    \partial_z\bar{p} &= -\bar{\rho}g,
\end{align}
\end{subequations}
where the radial pressure gradient provides the necessary centripetal force to balance the Coriolis and inertial effects of the background rotation, while the vertical gradient balances the static stratification. To maintain the steady-state assumption inherent in our flow state decomposition, we explicitly neglect the slow diffusion of the background flow, assuming it occurs on a timescale much longer than the dynamics of interest. This physical justification ensures that the background profile remains fixed, preventing non-physical drift, and implies that the viscous and diffusive operators in the governing equations act exclusively on the perturbation fields.

The governing equations of the system can be obtained by removing the background equilibrium from the Boussinesq equations. Noting that the advection of the linear background stratification simplifies to $(\bm{u}\cdot\grad)\bar{b} = -u'_z \bar{N}^2$, the governing equations are written in the vorticity form:
\begin{subequations}\label{eqn:boussinesq}
\begin{align}
    &\div\bm{u}' = 0, 
    \label{eqn:mass-conserv} \\
    &\partial_t \bm{u}'
    =  \bm{u}' \cross \bm{\omega}' - 2\Omega\dir{z}\cross\bm{u}' - b'\dir{z} -\left(\bm{u}'\cdot\grad\right)\bar{\bm{U}}-\left(\bar{\bm{U}}\cdot\grad\right)\bm{u}' + \nu\laplacian \bm{u}' -\grad \Pi , 
    \label{eqn:momentum} \\
    &\partial_t b' 
    = -(\bm{u}'\cdot\grad)b' + u'_z \bar{N}^2 -(\bar{\bm{U}}\cdot\grad)b'+ \kappa\laplacian b', 
    \label{eqn:density}
\end{align}
\end{subequations}
where $\bm{\omega}' = \curl\bm{u}'$, $\nu$ and $\kappa$ are the viscosity and thermal diffusivity of the fluid, respectively, and $2\Omega\dir{z}\times\bm{u}'$ in (\ref{eqn:momentum}) represents the Coriolis force acting on the disturbance velocity. Here, $\Pi$ is an effective pressure potential that absorbs the pressure ($p'/\rho_0$), the centrifugal force\footnote{The centrifugal acceleration associated with a constant background rotation along the vertical axis can be expressed as the gradient of a scalar potential.}, and the scalar potential associated with the vector identity $(\bm{u}'\cdot\grad)\bm{u}' = \grad(|\bm{u}'|^2/2) - \bm{u}' \cross \bm{\omega}'$.
It is important to emphasize that the two advective cross terms must share the same (advection or vorticity) form:
\begin{equation}
\begin{aligned}
    &\,-\left(\bm{u}'\cdot\grad\right)\bar{\bm{U}}-\left(\bar{\bm{U}}\cdot\grad\right)\bm{u}'\\
    =&\, \frac{1}{2}\left[\curl{(\bar{\bm{U}}\cross\bm{u}')}+\bm{u}'\cross\bar{\zeta}\dir{z}+\bar{\bm{U}}\cross\bm{\omega}'\right] + \frac{1}{2}\left[\curl{(\bm{u}'\cross\bar{\bm{U}})}+\bar{\bm{U}}\cross\bm{\omega}'+\bm{u}'\cross\bar{\zeta}\dir{z}\right] \\
    =&\, \bm{u}'\cross\bar{\zeta}\dir{z} + \bar{\bm{U}}\cross\bm{\omega}',
\end{aligned}
\end{equation}
which requires cancellation of the two curl operations in each cross term to achieve equality.

The Boussinesq equations (\ref{eqn:boussinesq}) can be non-dimensionalized using the reference density $\rho_0$, a characteristic velocity $U_0$, and a length scale $L_0$, where the choices of the latter two characteristic scales depend on the particular flows of interest. 
This gives a set of dimensionless parameters, which are the Rossby number ($Ro$), the Reynolds number ($Re$), and the horizontal Froude number ($Fr$):
\begin{subequations}
\begin{equation}
    Ro \equiv \frac{U_0}{2\Omega\cdot L_0},\quad
    Re \equiv \frac{U_0\cdot L_0}{\nu},\quad
    Fr \equiv \frac{U_0}{L_0 \cdot \bar{N}},
    \tag{\theequation a,b,c}
\end{equation}
\end{subequations}
where $Fr$ is the ratio between the advection timescale, $L_0/U_0$, and the characteristic timescale of the internal gravity waves, $1/\bar{N}$, and reflects the relative strength between the inertial forces and the buoyancy force.

\subsection{Conservation laws}
To establish a framework for the validation of our numerical schemes, we derive the conservation laws for the Boussinesq equations (\ref{eqn:boussinesq}). We define the disturbance kinetic energy (${E_\mathrm{K}}$) and available potential energy (${E_\mathrm{AP}}$) as:
\begin{subequations}
\begin{align}
    &{E_\mathrm{K}} \equiv \frac{1}{2}\int_V |\bm{u}'|^2 \dd{V}, \\
    &{E_\mathrm{AP}} \equiv \frac{1}{2\bar{N}^2}\int_V(b')^2dV.
\end{align}
\end{subequations}
The volume integration extends over the unbounded cylindrical domain ($0\leq r < \infty$), where we assume periodicity in the azimuthal ($\theta$) and axial ($z$) directions; this assumption of axial periodicity is justified by considering a domain length sufficient to isolate the dynamics of interest from boundary effects. Note that while the background buoyancy $\bar{b}(z)$ is linear and non-periodic, it is decoupled from (\ref{eqn:boussinesq}) and not affected by this assumption. 

The evolution equation for the kinetic energy is derived by taking the scalar product of the momentum equation (\ref{eqn:momentum}) with $\bm{u}'$ and integrating over the domain. We utilize the vorticity form of the nonlinear term, $\bm{u}' \times \bm{\omega}'$, which naturally vanishes upon contraction with $\bm{u}'$ due to vector orthogonality. Contributions from the Coriolis force similarly vanish, while all the flux terms integrate to zero via the divergence theorem (see Appendix~\ref{app:conservation_derivation} for detailed derivations). The evolution equation for the available potential energy is obtained by multiplying the buoyancy equation by $b'$ and integrating in the same way. The resulting energy budget is governed by:
\begin{subequations}\label{eqn:mechanical-energy-b'}
\begin{align}
    &\dv{t}E_\mathrm{K} = - \mathscr{E}_\mathrm{exc} - \mathscr{E}_\mathrm{shear} - \mathscr{E}_\mathrm{visc}, \label{eqn:KE-balance}\\
    &\dv{t}E_\mathrm{AP} = \mathscr{E}_\mathrm{exc} - \mathscr{E}_\mathrm{diff}, \label{eqn:APE-balance}\\
    &\dv{t} (E_\mathrm{K}+E_\mathrm{AP}) = - \mathscr{E}_\mathrm{shear} -\mathscr{E}_\mathrm{visc} - \mathscr{E}_\mathrm{diff}, \label{eqn:conservation-law}
\end{align}
\end{subequations}
where the energy exchange, shear production, and dissipation terms are defined as:
\begin{subequations}\label{eqn:energy-terms}
\begin{align}
    &\mathscr{E}_\mathrm{exc} = \int_V(b'\cdot u'_z) \dd{V}, \\
    &\mathscr{E}_\mathrm{shear} = \int_V\left(r\dv{r}\bar{\Omega}\right) u'_r u'_\theta \dd{V}, \\
    &\mathscr{E}_\mathrm{visc} = \nu \int_V |\grad \bm{u}'|^2 \dd{V}, \\
    &\mathscr{E}_\mathrm{diff} = \frac{\kappa}{\bar{N}^2}\int_V |\grad b'|^2 \dd{V}.
\end{align}
\end{subequations}
It is important to note that we employ the available potential energy, $E_\mathrm{AP}$, rather than the standard potential energy definition, $E_\mathrm{P} \equiv \int_V (z\cdot b') \dd{V}$, because the evolution equation of the latter is ill-defined in an axially periodic domain due to the non-vanishing boundary flux $\div{\left[b'z\bm{u}'\right]}$ that explicitly contains the non-periodic vertical coordinate $z$\footnote{For a truly unbounded domain in $z$ with decaying boundary conditions that are sufficiently fast, the surface flux vanishes, and corresponding conservation laws can be constructed.}. 

In the absence of diffusion, the total disturbance energy $E_\mathrm{K} + E_\mathrm{AP}$ evolves solely due to shear production term $\mathscr{E}_\mathrm{shear}$, which represents the energy exchange between the azimuthal shear of the background flow and the disturbance field. Note that ${E}_\mathrm{shear}$ only acts on $E_\mathrm{K}$ because the azimuthal shear is orthogonal to the direction of stratification. The buoyancy flux $\mathscr{E}_\mathrm{exc}$ appears with opposite signs in the $E_\mathrm{K}$ and $E_\mathrm{AP}$ equations, representing the reversible conversion between kinetic and potential energy via vertical motion. Consequently, if one considers the equivalent energy balance for the \textit{total} flow field (background plus disturbance), both interaction terms become internal to the system and do not appear, and the total energy of the full system is exactly conserved in the diffusion-free limit, as detailed in Appendix~\ref{app:conservation_derivation}. 


\subsection{Poloidal-toroidal decomposition}\label{sec:PT-decomp}
The governing Boussinesq equations (\ref{eqn:boussinesq}) are formulated in terms of the primitive perturbation variables $(p', b', \bm{u}')$. Because the velocity field is strictly solenoidal according to (\ref{eqn:mass-conserv}), we leverage a poloidal-toroidal decomposition \citep{chandrasekhar_1981,matsushima_1997,lee_2023,wang_2024} to enforce the divergence-free constraint and analytically eliminate the pressure variable.

By selecting the axial unit vector $\dir{z}$ as the reference vector, any generic solenoidal vector field $\bm{A}(r,\theta,z)$ can be expressed in terms of two scalar potentials:
\begin{equation}\label{eqn:PT-decomp}
    \bm{A} = \curl{(\psi\dir{z})} + \curl{\left[\curl{(\chi\dir{z})}\right]},
\end{equation}
where $\psi(r,\theta,z)$ and $\chi(r,\theta,z)$ are the toroidal and poloidal streamfunctions, respectively. Taking the axial component and the axial curl of this decomposition yields a pair of two-dimensional Poisson equations:
\begin{subequations}\label{eqn:pt-poisson}
\begin{align}
-\laplacian_\perp \chi &= \dir{z}\cdot\bm{A}, \\
-\laplacian_\perp \psi &= \dir{z}\cdot(\curl{\bm{A}}),
\end{align}
\end{subequations}
where the transverse Laplacian operator $\laplacian_\perp$ is defined as
\begin{equation}
    \laplacian_\perp \equiv \frac{1}{r}\pdv{}{r}r\pdv{}{r} - \frac{1}{r^2}\pdv[2]{}{\theta}.
\end{equation}
To uniquely determine $\psi$ and $\chi$ from a known vector field, one must invert these Poisson equations, which can be obtained by a two-dimensional convolution with the free-space Green's function, $G(r,\theta) = \ln{r}/(2\pi)$. As discussed by \citet{lee_2023}, this convolution is mathematically well-defined as long as the vector components decay algebraically or faster in the far field ($r\rightarrow\infty$), bounded by:
\begin{subequations}
\begin{equation}\label{eqn:far-conditions}
    A_r \sim \order{r^{-1-p_1}},\quad A_\theta \sim \order{r^{-1-p_2}},\quad A_z \sim \order{r^{-2-p_3}},
    \tag{\theequation a,b,c}
\end{equation} 
\end{subequations}
for constants $p_1, p_2, p_3 \geq 0$. If $\bm{A}$ represents a velocity field, these bounds guarantee that its kinetic energy density yields a finite total energy over the unbounded domain, which naturally satisfies our physical assumptions for the disturbance field. Furthermore, because the vector field is constructed purely from spatial derivatives of the poloidal-toroidal potentials, any constant offset is physically inconsequential. Hence, the remaining gauge freedom associated with the Laplacian inversion can be eliminated by imposing the following boundary conditions:
\begin{subequations}
\begin{align}
    \lim_{r\rightarrow\infty}\psi(r,\theta,z) &= 0, \\
    \lim_{r\rightarrow\infty}\chi(r,\theta,z) &= 0,
\end{align}
\end{subequations}
so a strictly linear and invertible projection operator, $\mathbb{P}_\mathrm{sol}$, can be defined between the space of compliant solenoidal vector fields and their poloidal-toroidal components:
\begin{equation}
\mathbb{P}_\mathrm{sol}\left[\bm{A}\right] \equiv  \begin{bmatrix} \psi \\ \chi \end{bmatrix}.
\end{equation}

In cylindrical coordinates, the spatial curl operators commute with both the full three-dimensional Laplacian ($\laplacian$) and the transverse Laplacian ($\laplacian_\perp$). It implies that the linear viscous diffusion operator in the governing Boussinesq equations can be applied directly to the scalar potentials rather than operating on the primitive velocity variables. Moreover, it allows us to extend our projection operator to a general projection filter, $\mathbb{P}$, by invoking the Helmholtz decomposition, which states that any arbitrary vector field $\bm{F}$ can be written as the sum of a strictly solenoidal field and an irrotational scalar gradient ($\bm{F} = \bm{A} + \grad{\Pi}$), so the extended projection operator isolates the poloidal-toroidal components of the solenoidal part:
\begin{equation}
    \mathbb{P}[\bm{F}] \equiv \mathbb{P}_\mathrm{sol}[\bm{A}].
\end{equation}
Note that while the poloidal potential $\chi$ for a strictly solenoidal field can be found directly from the axial component $A_z$, isolating the solenoidal part of a general, non-solenoidal field requires taking both the axial curl and the axial curl of the curl of $\bm{F}$ to annihilate the irrotational scalar field ($\curl{\grad{\Pi}} \equiv \bm{0}$). Exploiting the operator commutativity, this defines the generalized Poisson equations for the extended projection operator:
\begin{subequations}\label{eqn:pt-poisson-extended}
\begin{align}
    -\laplacian_\perp \psi &= \dir{z}\cdot(\curl{\bm{F}}), \\
    -\laplacian_\perp (-\laplacian \chi) &= \dir{z}\cdot(\curl{\curl{\bm{F}}}).
\end{align}
\end{subequations}
By solving (\ref{eqn:pt-poisson-extended}), the extended projection operator effectively filters out any scalar gradient ($\mathbb{P}[\grad{\Pi}] \equiv \bm{0}$) and extracts only the solenoidal component of $\bm{F}$.

Consequently, applying the extended projection operator to the momentum equations (\ref{eqn:momentum}) immediately drops the pressure gradient $\grad \Pi$. 
The definition of $\mathbb{P}$ then serves a dual purpose in our numerical framework. For traditional IMEX schemes (such as the Adams-Bashforth Crank-Nicolson scheme), $\mathbb{P}$ acts as a dimension-reduction technique, allowing the momentum equations for $\bm{u}'$ to be cast and evolved directly in the reduced $(\psi,\chi)$ space without the need to explicitly enforce the divergence-free condition. Meanwhile, in the semi-analytical ETD schemes to be formulated, diagonalizing the stiff linear operators is significantly more efficient in the primitive velocity space. In this context, the extended operator $\mathbb{P}$ functions as an exact, solenoidal projection filter. Rather than explicitly calculating the pressure field to update the velocity, we simply map an intermediate, potentially non-solenoidal velocity field to poloidal-toroidal space and back ($\bm{F}_\mathrm{sol} = \mathbb{P}_\mathrm{sol}^{-1}\left[\mathbb{P}[\bm{F}]\right]$). Because these projection inversions are evaluated purely algebraically in spectral space, this operation cleanly filters out the implicit pressure gradient and satisfies mass conservation without the need to formulate complex pressure boundary conditions typical of traditional fractional-step solvers.

\section{Numerical methods}\label{sec:numerical-methods}
To facilitate the development of our numerical schemes, we cast the continuous Boussinesq disturbance equations into a compact vector-operator notation. Defining the state vector in the primitive velocity basis as
\begin{equation}
    \bm{v} = [u'_r, u'_\theta, u'_z, b']^\intercal, 
\end{equation}
the governing equations can be written abstractly as:
\begin{equation}
    \partial_t \bm{v} = \mathcal{L}_{\mathrm{wave}}\bm{v} + \mathcal{L}_{\mathrm{diff}}\bm{v} + \mathcal{L}_{\mathrm{cross}}\bm{v} + \mathcal{N}(\bm{v}) - \bm{\Pi},
    \label{eqn:boussinesq-operator}
\end{equation}
subject to the solenoidal constraint $\div{\bm{u}'} = 0$. Here, the column vector $\bm{\Pi} = [\grad\Pi, 0]^\intercal$ absorbs all effective irrotational potentials, including the pressure perturbation, the centrifugal potential, and the scalar gradient derived from the vorticity form of the nonlinear terms. The constant-coefficient linear matrix operators governing the background rotation and stratification ($\mathcal{L}_{\mathrm{wave}}$) and the viscous/thermal diffusion ($\mathcal{L}_{\mathrm{diff}}$) are explicitly given by:
\begin{subequations}\label{eqn:operators-constant}
\begin{equation}
    \mathcal{L}_{\mathrm{wave}} =
    \begin{bmatrix}
        0 & 2\Omega & 0 & 0 \\
        -2\Omega & 0 & 0 & 0 \\
        0 & 0 & 0 & -1 \\
        0 & 0 & \bar{N}^2 & 0
    \end{bmatrix}, \quad
    \mathcal{L}_{\mathrm{diff}} =
    \begin{bmatrix}
        \nu\laplacian & 0 & 0 & 0 \\
        0 & \nu\laplacian & 0 & 0 \\
        0 & 0 & \nu\laplacian & 0 \\
        0 & 0 & 0 & \kappa\laplacian
    \end{bmatrix}.
    \tag{\theequation a,b}
\end{equation}
\end{subequations}
The remaining operators, representing the linear interactions with the steady background shear flow ($\mathcal{L}_{\mathrm{cross}}$) and the nonlinear self-advection ($\mathcal{N}$), act on the state vector to produce the following column vectors:
\begin{subequations}\label{eqn:operators-dynamic}
\begin{equation}
    \mathcal{L}_{\mathrm{cross}}\bm{v} =
    \begin{bmatrix}
        -(\bar{\bm{U}}\cdot\grad)\bm{u}' - (\bm{u}'\cdot\grad)\bar{\bm{U}} \\
        -(\bar{\bm{U}}\cdot\grad)b'
    \end{bmatrix}, \quad
    \mathcal{N}(\bm{v}) =
    \begin{bmatrix}
        \bm{u}'\times\bm{\omega}' \\
        -(\bm{u}'\cdot\grad)b'
    \end{bmatrix}.
    \tag{\theequation a,b}
\end{equation}
\end{subequations}

To integrate the system (\ref{eqn:boussinesq-operator}) in time, we rely on the method of fractional steps. It is important to clarify that throughout the numerical recipes presented in this section, superscript notations such as $t_{N+1/2}$ are used strictly to denote intermediate computational stages within a single time step, rather than evaluations of the flow field at actual half-time intervals. Furthermore, the integration time step $\Delta t$ is uniform, and active time-step adjustment mechanisms are not considered. Across all numerical schemes evaluated in this study, the numerical treatment of the diffusion and nonlinear terms remains strictly consistent: the viscous and thermal diffusion operator ($\mathcal{L}_{\mathrm{diff}}$) is integrated implicitly using the unconditionally stable Crank-Nicolson (CN) method, while the nonlinear self-advection ($\mathcal{N}$) is integrated explicitly using the second-order Adams-Bashforth (AB2) method. The fundamental difference between the schemes presented herein lies entirely in the specific numerical treatment of the linear wave ($\mathcal{L}_{\mathrm{wave}}$) and cross-term ($\mathcal{L}_{\mathrm{cross}}$) operators. 

To provide a clear roadmap of the numerical implementation, this section is structured as follows. \S\ref{sec:radial-discretization} presents the spectral expansion. \S\ref{sec:baseline-ab-cn} details the numerical stiffness associated with standard IMEX methods. \S\ref{sec:basic-etd} formulates the semi-analytical ETD method, using the basic $\mathcal{L}_{\mathrm{wave}}$ operator as an illustrating example. We also detail the analytical handling of the pressure gradient and the resolution of zero-eigenvalue singularities. \S\ref{sec:advanced-etd} presents the full ETD scheme, explicitly deriving the diagonalization of the linear cross terms. Finally, \S\ref{sec:stabilization} discusses numerical stabilization, including hyperviscosity and wave absorption at domain boundaries.

\subsection{Spectral expansion}\label{sec:radial-discretization}
The unbounded cylindrical domain is expanded using Fourier series in the periodic azimuthal and axial directions, and mapped associated Legendre polynomials in the radial direction \citep{matsushima_1997}. Specifically,  The semi-infinite radial domain ($0 \leq r < \infty$) is mapped onto a finite domain $\mu \in [-1, 1]$ via the algebraic transformation:
\begin{equation}\label{eqn:algebraic-mapping}
    \mu(r) = \frac{r^2 - L^2}{r^2 + L^2},
\end{equation}
where $L > 0$ is a tunable mapping parameter. The radial basis is then constructed from the standard associated Legendre polynomials, $P^m_n(\mu)$:
\begin{equation}
    P^m_{Ln}(r) \equiv P^m_n(\mu(r)),
\end{equation}
and an arbitrary scalar field $\phi$ (representing any component of $\bm{v}$) can thus be expanded as
\begin{equation}\label{eqn:spectral-projection}
    \phi(r,\theta,z,t) = \sum_{m,k=-\infty}^{\infty}\sum_{n=|m|}^\infty \tilde{\phi}_{mnk}(t)\, P^m_{Ln}(r)\, e^{\iu(m\theta+kz)},
\end{equation}
where $m$ and $k$ are the azimuthal and axial wavenumbers, and $n$ is the degree of the associated Legendre polynomials. For any real scalar field, $\phi \in \mathbb{R}^3$, the reality condition further requires complex conjugate relations between $(m,k)$ and $(-m,-k)$ expansions, so we only need to keep half of the wavenumbers, e.g., $m\geq 0$. 

The calculation of the spectral coefficients $\tilde{\phi}_{mnk}(t)$ is based on the orthogonality condition:
\begin{equation}
    \int_0^\infty P^m_{Ln}(r) P^m_{Ln'}(r) w(r) \dd{r} = \delta_{nn'} \left(N^m_n\right)^2,
\end{equation}
where $\left(N^m_n\right)^2 = {2(n+|m|)!}/{((2n+1)(n-|m|)!)}$ is the normalization constant and $w(r) = 4L^2 r / (r^2 + L^2)^2$ is the mapped weight function. By exploiting this orthogonality relation, the spectral coefficients are extracted by integrating the product of the physical field $\hat{\phi}_{mk}(r,t)$, the basis function $P^m_{Ln}(r)$, and the weight function $w(r)$ over the radial domain. In practice, because this continuous integral over $[0, \infty)$ maps perfectly back to the bounded interval $\mu \in [-1, 1]$, the numerical integration can be evaluated efficiently using standard Gauss-Legendre quadratures defined on the $\mu$ grid. 

The mapped associated Legendre basis is highly advantageous for unbounded cylindrical domains for several theoretical and computational reasons. To begin with, the analytical mapping allocates exactly half of the radial collocation points to the inner region $0 \leq r \leq L$, yielding naturally high numerical resolution where vortex dynamics and shear gradients are typically most prominent. This localized resolution tuning was effectively exploited by \citet{lee_2023} in their linear stability analysis of incompressible columnar vortices. Additionally, the basis functions behave as $C^\infty$ at $r=0$, guaranteeing the requisite analyticity conditions at the coordinate singularity. In the far field ($r \to \infty$, corresponding to $\mu \to 1$), the mapped polynomials decay as $\order{r^{-|m|}}$, so, for any $C^\infty$ function that decays algebraically or exponentially as $r \to \infty$, the mapped associated Legendre polynomials natively form a Galerkin basis set that achieves exponential spectral convergence \citep{matsushima_1997,lee_2023}.

It is important to mention that, when performing the poloidal-toroidal decomposition, expanding the scalar streamfunctions ($\psi$ and $\chi$) directly in the mapped associated Legendre basis requires the inclusion of a supplemental logarithmic term, $P_l(r) \equiv \ln\left({(L^2 + r^2)}/{(2L^2)}\right)$,  
which is necessary to account for the $\order{r^{-1}}$ far-field asymptotic behavior of the azimuthal velocity field ($u_\theta$) and the mean axial components of the velocity and vorticity fields \citep{matsushima_1997}. Fortunately, taking the derivative of this logarithmic term maps it perfectly back into the standard polynomial space: $r\dv{r} P_l = P^0_{L0} + P^0_{L1}$, so the primitive velocity ($\bm{u}'$) and vorticity ($\bm{\omega}'$) fields themselves are expanded completely and exactly using solely the standard $P^m_{Ln}(r)$ basis without any logarithmic corrections. Moreover, as detailed in Appendix~\ref{app:math-identities}, the projected logarithmic coefficients $\psi_l$ for the nonlinear term $\bm{u}'\times\bm{\omega}'$ and the linear wave term $-(2\Omega\dir{z}\times\bm{u}'+b'\dir{z})$ naturally evaluate to exactly zero, ensuring that it does not dynamically complicate the time evolution equations.

We shall also note that the density perturbation $b'$ is subjected to a specialized physical constraint. Because the prescribed background stratification $\bar{b}(z)$ is horizontally uniform and dynamically rigid by definition, it cannot dynamically redistribute or absorb a net horizontal mass deficit. Therefore, any physically realizable localized perturbation must inherently restrict its own total mass anomaly to be finite within the unbounded domain, implying that its axisymmetric ($m=0$) modes must decay rapidly, at least as fast as $\order{r^{-4}}$, as $r\rightarrow\infty$. In the mapped domain, this suggests that the axisymmetric modes must be perfectly divisible by $(1-\mu)^2$, which mathematically translates to:
\begin{equation}\label{eqn:buoyancy-constraint}
    \sum_{n=0}^\infty \tilde{b}'_{0nk} = 0 \quad \text{and} \quad \sum_{n=0}^\infty \tilde{b}'_{0nk}\cdot\frac{n(n+1)}{2} = 0 \quad \text{for all } k.
\end{equation}
Since the governing equations and the spectral spatial operators conserve the domain-integrated mass anomaly, the spectral constraints above only need to be satisfied by the initial conditions, and the numerical scheme will preserve this asymptotic decay without requiring explicit filtering during time integration.

With the complete spectral basis, the governing equations (\ref{eqn:boussinesq-operator}) are integrated following a pseudo-spectral approach, which requires transitioning among three distinct computational spaces:
\begin{itemize}
    \item \textbf{FFF (Function-Function-Function) space:} The fully spectral domain where spatial variables are expanded by (\ref{eqn:spectral-projection}). Linear operators with constant coefficients (such as diffusion) are evaluated here as algebraic operations.
    \item \textbf{PPP (Physical-Physical-Physical) space:} The fully physical domain evaluated at discrete spatial grid points.
    \item \textbf{PFF (Physical-Function-Function) space:} A hybrid space where the azimuthal and axial dimensions remain in Fourier space, but the radial dimension is evaluated at physical collocation points to accommodate the radial dependencies of $\mathcal{L}_\mathrm{cross}$.
\end{itemize}
To evaluate the nonlinear advective product $\mathcal{N}(\bm{v})$, we bypass computationally expensive convolution integrals in the spectral space by transforming the state vector to the PPP space, where direct point-by-point multiplication is performed. The result is subsequently transformed back to the FFF space before the linear operators are applied. While highly efficient, this pointwise evaluation in the physical domain, combined with spectral truncation, introduces aliasing errors, which will be addressed in \S\ref{sec:stabilization}.

Beyond evaluating nonlinearities, our spectral representation offers significant computational advantages for the linear physics, primarily through the sparsity and highly structured nature of the resulting matrices. For instance, applying the transverse Laplacian $\laplacian_\perp$ to the basis function $P^m_{Ln}$ yields the analytical relation $ \laplacian_{\perp}|_{m, n} = (4n(n+1)L^2)/(L^2+r^2)^2$.
This property, along with the other operational recurrence relations documented by \citet{matsushima_1997}, ensures that radial differential operators map natively to narrow-banded matrices, drastically reducing both the computational complexity and the memory footprint of the numerical solver. 

Furthermore, because the azimuthal and axial dimensions are expanded in the orthogonal Fourier basis $e^{\iu(m\theta+kz)}$, the three-dimensional linear operators (e.g., $\mathcal{L}_{\mathrm{wave}}$, $\mathcal{L}_{\mathrm{cross}}$, and $\mathcal{L}_{\mathrm{diff}}$) decouple entirely with respect to $m$ and $k$. Consequently, they decompose into a set of independent, banded block matrices for each $(m,k)$ pair, which is optimally suited for massive parallelization. In our implementation, the spectral $(m,k)$ modes are distributed across independent computing processors via a 2D pencil decomposition using the Message Passing Interface (MPI) \citep[see, for example,][]{dalcin_2019, MLEGS}. This data distribution strategy allows the linear operators to be evaluated simultaneously across all processors without any inter-process communication. 

In fact, cross-processor communication is incurred exclusively during the computational space conversions between the FFF and PPP spaces required for the nonlinear products. Because Fast Fourier Transforms (FFTs) necessitate contiguous memory access to all data points along a given spatial dimension, the global data array must be transposed and realigned along the two pencil decomposition directions using MPI all-to-all communications. While the linear physics scale perfectly, these global memory transposes become the primary bandwidth bottleneck for large-scale, multi-node deployments.

\subsection{Stiffness associated with the explicit treatment of the background terms}\label{sec:baseline-ab-cn}
A standard and widely utilized approach to time integration relies on treating the linear operators $\mathcal{L}_{\text{wave}}$ and $\mathcal{L}_{\text{cross}}$ explicitly alongside the nonlinear terms. The primary motivation for this formulation is its ability to operate primarily within the reduced poloidal-toroidal state space, $\bm{V} = [\psi, \chi, b']^\intercal$. By converting the primitive velocity components to scalar potentials, the number of evolving variables is reduced from four to three.

Conveniently, because the curl operators defining the poloidal-toroidal decomposition commute perfectly with the three-dimensional Laplacian ($\mathbb{P}\laplacian\bm{u}' = \laplacian\mathbb{P}[\bm{u}']$), the implicit diffusion operator $\mathcal{L}_{\text{diff}}$ can be evaluated directly on the poloidal-toroidal basis. Furthermore, applying the forward poloidal-toroidal projection ($\mathbb{P}$) natively extracts the solenoidal component of any vector field, thereby inherently annihilating the irrotational potential field ($\mathbb{P}[\grad\Pi] \equiv \bm{0}$) without requiring a dedicated pressure Poisson solver. The numerical scheme proceeds as follows:
\begin{recipe}[Explicit AB2-CN scheme in poloidal-toroidal space]
\leavevmode
\normalfont
\begin{enumerate}
    \item \textbf{Explicit Forcing and Projection:} Given the spectral poloidal-toroidal state $\bm{V}^{t_N}$, reconstruct the primitive variables:
        \begin{equation}
            \bm{v}^{t_N} = \mathbb{P}_\mathrm{sol}^{-1}[\bm{V}^{t_N}].
        \end{equation} 
    Evaluate the explicit forcing vector in the physical (PPP) space and project to poloidal-toroidal components in spectral (FFF) space:
        \begin{equation}
            \bm{F}^{t_N} = \mathbb{P}\left[\mathcal{L}_{\text{wave}}\bm{v}^{t_N} + \mathcal{L}_{\text{cross}}\bm{v}^{t_N} + \mathcal{N}(\bm{v}^{t_N})\right].
        \end{equation}
    \item \textbf{AB2 Extrapolation:} Advance the projected explicit forcing using the historical Adams-Bashforth weights to form an intermediate, undiffused state:
        \begin{equation}
            \bm{V}^{t_{N+1/2}} = \bm{V}^{t_N} + \Delta t \left( \frac{3}{2}\bm{F}^{t_N} - \frac{1}{2}\bm{F}^{t_{N-1}} \right).
        \end{equation}
    \item \textbf{Implicit Diffusion with CN:} Solve the resulting uncoupled, implicit Helmholtz equations for the diffused poloidal-toroidal state to finalize the time step:
        \begin{equation}
            \bm{V}^{t_{N+1}} = \left(\textrm{diag}^{-1}(\nu,\nu,\kappa)\frac{2}{\Delta t} - \laplacian\right)^{-1} \left(\textrm{diag}^{-1}(\nu,\nu,\kappa)\frac{2}{\Delta t}\bm{V}^{t_{N+1/2}} + \laplacian\bm{V}^{t_{N}}\right).
        \end{equation}
\end{enumerate}
\end{recipe}

Because this scheme operates natively in the poloidal-toroidal space, it features a low memory footprint and is computationally efficient per time step. The commutativity of the diffusion operator ensures that the implicit solve natively preserves the divergence-free condition, meaning the pressure treatment is strictly isolated to the explicit projection step and handled by the poloidal-toroidal projection. This scheme is notably robust and useful for physical scenarios characterized by a purely hydrostatic background state under low background rotation and stratification.

However, the explicit treatment of $\mathcal{L}_{\mathrm{wave}}$ and $\mathcal{L}_{\mathrm{cross}}$ introduces a severe operational limitation for rapidly rotating or strongly stratified flows. By inspection, the matrix operator $\mathcal{L}_{\mathrm{wave}}$ as given in (\ref{eqn:operators-constant}a) exhibits a strict $2\times 2$ block-diagonal structure, whose eigenvalues are trivial to determine: $\lambda_{\mathrm{wave}} = \pm \iu 2\Omega$ and $\pm \iu \bar{N}$. These purely imaginary eigenvalues correspond precisely to the physical frequencies of the epicyclic (inertial) oscillations and the internal gravity waves. Treating such restorative, high-frequency physics explicitly using the Adams-Bashforth method strictly bounds the allowable time step to maintain absolute numerical stability. Specifically, to keep these purely imaginary eigenvalues marginally within the numerical stability region, the integration time step is severely bottlenecked by the fastest linear waves\footnote{We assume a stably stratified background ($\bar{N}^2 > 0$) in this stiffness analysis, which yields the purely imaginary eigenvalues responsible for the severe oscillatory constraint. It is worth noting, however, that the semi-analytical exponential time differencing scheme developed in subsequent sections remains mathematically valid and strictly accurate for unstable stratifications ($\bar{N}^2 < 0$), as it naturally and exactly integrates the resulting real exponential growth factors.}: $\Delta t \lesssim \mathcal{O}(1/\max(2\Omega, \bar{N}))$. 

Furthermore, when a background flow possesses strong azimuthal shear, advection dominates the cross-term operator $\mathcal{L}_{\mathrm{cross}}$, yielding eigenvalues that scale with the local angular velocity and imposing an additional advective constraint on the time step. While the particular stiffness arising from these radially dependent cross terms will be formally elaborated and resolved in \S\ref{sec:advanced-etd}, its presence here compounds the problem. Consequently, for flows where the background rotation, stratification, or shear is significantly faster than the timescale of the nonlinear, small-scale disturbances, this explicit formulation forces the numerical scheme to take prohibitively small time steps, rendering long-term integration impractical.

We must emphasize that, while it may seem intuitive to bypass these explicit stability limits by absorbing $\mathcal{L}_{\text{wave}}$ and $\mathcal{L}_{\text{cross}}$ into the implicit Crank-Nicolson solve alongside diffusion, doing so introduces severe numerical artifacts. The Crank-Nicolson method approximates the exact exponential time evolution using a $(1,1)$ Padé approximant. For purely imaginary eigenvalues, which perfectly characterize the restorative, oscillatory nature of inertial-gravity and epicyclic waves, this approximation is unconditionally stable, yielding an amplification factor with a modulus of exactly $1$. However, at large Courant numbers, the Padé approximant suffers from severe phase-speed retardation. This numerical dispersion causes high-frequency waves to artificially slow down and steepen, generating spurious ``ringing'' oscillations. In rotating and stratified shear flows, these non-physical dispersive artifacts phenomenologically masquerade as genuine vortex instabilities. Therefore, an implicit CN treatment of the wave and cross terms cannot be trusted, which motivates the need for a semi-analytical integration method.

\subsection{Semi-analytical time-domain integration with exponential time differencing}\label{sec:basic-etd}
To resolve the numerical stiffness without artificially retarding phase speeds or inducing numerical dispersion, we deploy an exponential time differencing scheme, which exponentiates the stiff linear operators and integrates the linear wave and advective physics analytically over the time step. 

Unlike the AB2-CN scheme, the ETD formulation operates primarily in the primitive velocity space ($\bm{v}$). This is because exponentiating the fully coupled linear operators in the primitive form allows for a clean and straightforward diagonalization, whereas the poloidal-toroidal form is not easily tractable. To establish the mathematical foundation of our approach and formulate the general recipe, we first derive a basic ETD scheme that isolates and exponentiates solely the constant-coefficient wave operator\footnote{This basic ETD scheme also adequately handles a hydrostatic background.}.

To maintain strict notational consistency for the matrix exponentials and primitive-form vectors, we let:
\begin{subequations}
\begin{equation}
    \mathsfbi{L} \equiv \mathcal{L}_{\mathrm{wave}},\quad \bm{f}(\bm{v}) = \mathcal{L}_{\mathrm{cross}}\bm{v} + \mathcal{N}(\bm{v}), \quad \bm{g}(\bm{v}) = \mathcal{L}_{\mathrm{diff}}\bm{v}(t) - \bm{\Pi}(t).
    \tag{\theequation a,b,c}
\end{equation}    
\end{subequations}
Multiplying the continuous governing equation by the matrix integrating factor $e^{-\mathsfbi{L}t}$ gives:
\begin{equation}
    \partial_t \left(e^{-\mathsfbi{L}t}\bm{v}\right) = e^{-\mathsfbi{L}t}\left(\bm{f}(t) + \bm{g}(t)\right),
\end{equation}
which can be integrated exactly over a single time step from $t_N$ to $t_{N+1}$:
\begin{equation}
\begin{aligned}
    \bm{v}^{t_{N+1}} &= e^{\mathsfbi{L}\Delta t}\bm{v}^{t_N} + \int_0^{\Delta t} e^{\mathsfbi{L}(\Delta t - \tau)} \left( \bm{f}(t_N+\tau) + \bm{g}(t_N+\tau) \right) \dd{\tau} \\
    &= \mathsfbi{E}\bm{v}^{t_N} + \mathsfbi{NL} \left( \frac{3}{2}\bm{f}^{t_N} - \frac{1}{2}\bm{f}^{t_{N-1}} + \frac{1}{2}\bm{g}^{t_N} + \frac{1}{2}\bm{g}^{t_{N+1}} \right) + \order{\Delta t^3},
    \label{eqn:etd-approx-intermediate}
\end{aligned}
\end{equation}
where the discrete integration operators are defined as:
\begin{subequations}
\begin{equation}
    \mathsfbi{E} \equiv e^{\mathsfbi{L}\Delta t},\quad 
    \mathsfbi{NL} \equiv \int_0^{\Delta t} e^{\mathsfbi{L}(\Delta t - \tau)} \dd{\tau} = \mathsfbi{L}^{-1}\left(e^{\mathsfbi{L}\Delta t}-\mathsfbi{I}\right).
    \tag{\theequation a,b}
\end{equation}    
\end{subequations}
Note here that the explicit forcing, the diffusive terms, and the irrotational potential all continuously evolve over the integration interval $\tau \in [0, \Delta t]$, and their integrals are approximated as constants to be factored out of the integration operator. Specifically, the explicit term $\bm{f}$ is extrapolated using the historical second-order Adams-Bashforth weights, while the implicit diffusion and potential fields $\bm{g}$ are approximated using a Crank-Nicolson temporal average. 

The right-hand side of (\ref{eqn:etd-approx-intermediate}) contains the unknown future state $\bm{g}^{t_{N+1}}$. To isolate it without corrupting the second-order accuracy, we exploit the Taylor series expansions of the two ETD operators:
\begin{subequations}
\begin{align}
    \mathsfbi{E} &= \mathsfbi{I} + \Delta t \mathsfbi{L} + \frac{{\Delta t}^2}{2} \mathsfbi{L}^2 + \order{\Delta t^3}, \\
    \mathsfbi{NL} &= \Delta t \mathsfbi{I} + \frac{{\Delta t}^2}{2} \mathsfbi{L} + \order{\Delta t^3}.
\end{align}
\end{subequations}
By expanding the grouped implicit terms from (\ref{eqn:etd-approx-intermediate}), one can show that applying the $\mathsfbi{NL}$ operator to the CN average is mathematically equivalent, up to $\order{\Delta t^3}$ local truncation error, to applying the exponential operator directly to the historical implicit state:
\begin{equation}
    \mathsfbi{NL}\left(\frac{1}{2}\bm{g}^{t_N} + \frac{1}{2}\bm{g}^{t_{N+1}}\right) = \mathsfbi{E}\tilde{\bm{g}}^{t_N} + \tilde{\bm{g}}^{t_{N+1}} + \order{\Delta t^3},
\end{equation}
where we define the time-scaled implicit quantity as $\tilde{\bm{g}} \equiv (\Delta t/2)\bm{g}$ and make use of the expansion $\tilde{\bm{g}}^{t_{N+1}} = \tilde{\bm{g}}^{t_{N}} + \order{\Delta t^2}$. Because the local truncation error is $\order{\Delta t^3}$, this substitution preserves the second-order global accuracy. Substituting this consistent approximation back into our numerical formulation and expanding $\tilde{\bm{g}}$ separates the known and unknown states:
\begin{equation}
    \left(\mathsfbi{I} - \frac{\Delta t}{2}\mathcal{L}_{\mathrm{diff}}\right)\bm{v}^{t_{N+1}} + \tilde{\bm{\Pi}}^{t_{N+1}} = \mathsfbi{E}\left(\bm{v}^{t_N} + \frac{\Delta t}{2}\mathcal{L}_{\mathrm{diff}}\bm{v}^{t_N} - \tilde{\bm{\Pi}}^{t_N}\right) + \mathsfbi{NL}\left(\frac{3}{2}\bm{f}^{t_N} - \frac{1}{2}\bm{f}^{t_{N-1}}\right),
\label{eqn:etd-final-algebraic}
\end{equation}
where $\tilde{\bm{\Pi}} \equiv (\Delta t/2)\bm{\Pi}$ is the scaled irrotational potential vector. The right-hand side of equation~(\ref{eqn:etd-final-algebraic}) depends strictly on known quantities, forming an intermediate state $\bm{v}^{t_{N+1/2}}$. Applying the continuous poloidal-toroidal projection $\mathbb{P}$ to both sides natively annihilates the updated potential $\tilde{\bm{\Pi}}^{t_{N+1}}$. Because $\mathbb{P}$ and $\mathcal{L}_{\mathrm{diff}}$ commute, this algebraic reduction directly returns the uncoupled, implicit Helmholtz equations for the diffused poloidal-toroidal state.

To compute the exact matrix exponentials, the $4\times 4$ constant-coefficient operator $\mathsfbi{L} = \mathcal{L}_\mathrm{wave}$ is diagonalized as 
\begin{equation}
    \mathsfbi{L} = \mathsfbi{S}\mathsfbi{J}\mathsfbi{S}^{-1},
\end{equation}
where $\mathsfbi{J}$ is the diagonal matrix of eigenvalues, representing the inertial and internal gravity wave frequencies, and $\mathsfbi{S}$ and $\mathsfbi{S}^{-1}$ are the corresponding eigenvector matrix and its analytical inverse:
\begin{subequations}
\begin{align}
    \mathsfbi{J} &= \mathrm{diag}\left(\mathrm{i}2\Omega,\, -\mathrm{i}2\Omega,\, \mathrm{i}\bar{N},\, -\mathrm{i}\bar{N}\right), \\
    \mathsfbi{S} &=
    \begin{bmatrix}
        1 & 1 & 0 & 0 \\
        \mathrm{i} & -\mathrm{i} & 0 & 0 \\
        0 & 0 & 1 & 1 \\
        0 & 0 & -\mathrm{i}\bar{N} & \mathrm{i}\bar{N}
    \end{bmatrix}, \quad\quad
    \mathsfbi{S}^{-1} = \frac{1}{2}
    \begin{bmatrix}
        1 & -\mathrm{i} & 0 & 0 \\
        1 & \mathrm{i} & 0 & 0 \\
        0 & 0 & 1 & \mathrm{i}/\bar{N} \\
        0 & 0 & 1 & -\mathrm{i}/\bar{N}
    \end{bmatrix}.
\end{align}
\end{subequations}
The discrete integration operators can then be directly evaluated as:
\begin{subequations}
\begin{equation}
    \mathsfbi{E} = \mathsfbi{S}e^{\mathsfbi{J}\Delta t}\mathsfbi{S}^{-1},\quad
    \mathsfbi{NL} = \mathsfbi{S}\mathsfbi{J}^{-1}\left(e^{\mathsfbi{J}\Delta t}-\mathsfbi{I}\right)\mathsfbi{S}^{-1},
    \tag{\theequation a,b}
\end{equation}
\end{subequations}
which remain constant across time steps and can be precomputed during initialization. 

An important technical detail associated with the numerical evaluation of the $\mathsfbi{NL}$ operator is that, whenever $\mathsfbi{J}$ contains zero eigenvalues (e.g., when the background rotation is absent, $\Omega=0$, causing $\mathsfbi{J}^{-1}$ to become undefined), a singularity arises and must be addressed. This indeterminate form is strictly mathematical, not physical, and is resolved by applying L'Hôpital's rule for the corresponding diagonal components, denoted by $\lambda$:
\begin{equation}
    \lim_{\lambda\rightarrow 0}{\frac{e^{\lambda\Delta t}-1}{\lambda}} = \Delta t.
    \label{eqn:etd-eigval-limit}
\end{equation}
Consequently, any zero-eigenvalue of the diagonalized linear operator smoothly asymptotes to $\Delta t$, effectively reducing the ETD integration to a standard AB2 explicit update for that specific component (for the non-rotating case, the horizontal velocity components are reduced to AB2 for time integration).

Likewise, a critical physical singularity arises for the $z$-invariant modes. Consider the $k=0$ wavenumber of the disturbance field: $\bm{u}'_{m0} \equiv \bm{u}'_{m0}(r,\theta)$. The solenoidal condition (\ref{eqn:mass-conserv}) is linear and suggests that it can be constructed from a set of poloidal-toroidal potential functions: $\bm{u}'_{m0} = (\partial_\theta\psi'_{m0}/r)\dir{r} - \partial_r\psi'_{m0}\hat{\bm{\theta}}-\laplacian_\perp\chi'_{m0}\dir{z}$. The result of $\mathcal{L}_\mathrm{wave}$ applied to $\bm{u}'_{m0}$ is then evaluated to be $2\Omega \partial_r\psi'_{m0} \dir{r} + (2\Omega\partial_\theta\psi'_{m0}/r)\hat{\bm{\theta}} - b'_{m0}\dir{z}$, whose divergence is given by $2\Omega \laplacian \psi_{m0}$. Hence, it is instantaneously balanced by the two-dimensional pressure field ($\Pi_{m0} = 2\Omega \psi_{m0}$), leaving only the axial component, so the Coriolis force exerts exactly zero net dynamic effect on the physical evolution of the $k=0$ modes. If the standard $\mathsfbi{E}$ and $\mathsfbi{NL}$ operators are blindly applied to these modes, the scheme will unphysically rotate them, corrupting the kinetic energy of the flow. To enforce the correct physics, the discrete horizontal ETD operators for all $k=0$ modes must be manually overridden to match their non-rotating limits: $\mathsfbi{E}_{k=0} \to \mathsfbi{E}_{\Omega=0}$ and $\mathsfbi{NL}_{k=0} \to \mathsfbi{NL}_{\Omega=0}$.

With the operators explicitly defined and the constraints satisfied, the basic ETD time-stepping algorithm proceeds as follows\footnote{Richardson extrapolation can be applied to initiate the time-stepping, which consists of two half steps and one full step, each following a first-order exponential time differencing scheme that uses forward Euler method to extract the nonlinear and cross terms, and backward Euler method for the pressure term and the diffusion term.}:
\begin{recipe}[ETD scheme in primitive space]
\leavevmode
\normalfont
\begin{enumerate}
    \item \textbf{Explicit Forcing and Filtering:} Given the primitive state $\bm{v}^{t_N}$, compute the explicit forcing $\bm{f}^{t_N}$ in the physical (PPP) space, and project the explicit terms to poloidal-toroidal space and back to enforce strict incompressibility prior to time integration for numerical stability:
        \begin{equation}\label{eqn:etd-explicit-forcing}
            \bm{f}_\mathrm{sol}^{t_N} = \mathbb{P}_\mathrm{sol}^{-1}\left[\mathbb{P}[\mathcal{L}_{\mathrm{cross}}\bm{v}^{t_N} + \mathcal{N}(\bm{v}^{t_N})]\right].
        \end{equation}
    \item \textbf{ETD Advancement:} Apply the exact matrix exponentials to the historical primitive state and the AB2-extrapolated explicit forcing to obtain an intermediate, undiffused state:
        \begin{equation}
            \bm{v}^{t_{N+1/2}} = \mathsfbi{E}\left(\bm{v}^{t_{N}} + \frac{\Delta t}{2}\mathcal{L}_{\mathrm{diff}}\bm{v}^{t_N} - \tilde{\bm{\Pi}}^{t_{N}}\right) + \mathsfbi{NL} \left( \frac{3}{2}\bm{f}_\mathrm{sol}^{t_N} - \frac{1}{2}\bm{f}_\mathrm{sol}^{t_{N-1}} \right).
        \end{equation}
    \item \textbf{Implicit Diffusion:} Apply the forward poloidal-toroidal projection $\mathbb{P}$ to algebraically drop the potential gradient $\tilde{\bm{\Pi}}^{t_{N+1}}$. Solve the resulting uncoupled, implicit Helmholtz equations in the spectral (FFF) space for the diffused poloidal-toroidal state to finalize the explicit physics:
        \begin{equation}
            \bm{V}^{t_{N+1}} = \left(\mathsfbi{I} - \frac{\Delta t}{2}\mathcal{L}_{\mathrm{diff}}\right)^{-1} \mathbb{P}[\bm{v}^{t_{N+1/2}}].
        \end{equation}
        The final primitive velocity is subsequently reconstructed via 
        \begin{equation}
            \bm{v}^{t_{N+1}} = \mathbb{P}_\mathrm{sol}^{-1}[\bm{V}^{t_{N+1}}].
        \end{equation}
    \item \textbf{Pressure Extraction:} Calculate the updated irrotational potential using the residual of the intermediate state, to be stored for the subsequent time step:
        \begin{equation}
            \tilde{\bm{\Pi}}^{t_{N+1}} = \bm{v}^{t_{N+1/2}} - \left(\mathsfbi{I} - \frac{\Delta t}{2}\mathcal{L}_{\mathrm{diff}}\right)\bm{v}^{t_{N+1}}.
        \end{equation}
\end{enumerate}
\end{recipe}

It is worth mentioning that the poloidal-toroidal projection in the explicit forcing evaluation (\ref{eqn:etd-explicit-forcing}) is an optional treatment to improve the numerical stability. In essence, it decomposes the pressure term into two distinct components: the first instantaneously addresses velocity changes caused by nonlinear interactions, while the second accounts for changes arising from the explicit time integration of the linear term. While this additional step neither increases nor reduces the order of the local truncation error of the numerical scheme, it ensures that any field that is fed into the time integration operators of the ETD schemes is divergence-free, which improves the overall stability of the code.

\subsection{Diagonalization of the background shear}\label{sec:advanced-etd}
While the basic ETD scheme formulated above stabilizes the restorative wave forces, evaluating the background shear cross terms explicitly via the Adams-Bashforth method remains impractically stiff for strongly sheared flows. To achieve robust simulation speeds, the analytical integration must be expanded to encompass these shear interactions alongside the wave operator. However, to incorporate the cross terms into the exact matrix exponential, we must carefully select their mathematical formulation, as it directly dictates the computational tractability of the resulting scheme. 

If $\mathcal{L}_{\mathrm{cross}}$ is expressed in the vorticity form, $\mathcal{L}_{\mathrm{cross}}(\bm{v}) = \bar{\bm{U}} \times \bm{\omega}' + \bm{u}' \times \bar{\zeta}(r)\dir{z}$, it isolates two distinct physical mechanisms: the macroscopic sweeping and stretching of the perturbation vortices ($\bar{\bm{U}} \times \bm{\omega}'$) and the local Coriolis-like restoring force driven by the background vorticity ($\bm{u}' \times \bar{\zeta}\dir{z}$). While physically insightful, the curl operator required to obtain $\bm{\omega}'$ inextricably couples the different velocity components within the vorticity stretching term. Exponentiating this differential operator would require the inversion of a massive, globally coupled matrix system, rendering the exact integration computationally prohibitive. Alternatively, one might consider a hybrid scheme to circumvent this structural issue, where the purely algebraic term $-\bar{\zeta}\dir{z}\times\bm{u}'$ is analytically diagonalized alongside the wave operator, and the problematic differential term $\bar{\bm{U}} \times \bm{\omega}'$ is absorbed into the explicitly evaluated forcing vector $\bm{f}(\bm{v})$. However, our numerical experiments showed this partial ETD implementation inevitably suffers from numerical blow-up at moderate time steps, indicating that the scheme is still critically stiff due to the explicit treatment of the remaining cross term.

Casting the cross terms in their primitive advective form, however, resolves this structural limitation: $\mathcal{L}_{\mathrm{cross}} = -(\bar{\bm{U}} \cdot \grad)\bm{u}' - (\bm{u}' \cdot \grad)\bar{\bm{U}}$. Because the background flow is strictly azimuthal ($\bar{\bm{U}} = r\bar{\Omega}(r)\hat{\bm{\theta}}$), evaluating these directional derivatives of the perturbation vector in cylindrical coordinates generates geometric off-diagonal terms alongside the advective diagonal $-\iu m \bar{\Omega}(r)$, all of which do not contain the radial derivatives. The fully coupled linear operator $\mathsfbi{L} = \mathcal{L}_{\mathrm{wave}} + \mathcal{L}_{\mathrm{cross}}$ is thus strictly algebraic in $r$. This guarantees that in the hybrid PFF space, $\mathsfbi{L}$ forms an independent, $4\times 4$ block-diagonal matrix at each radial collocation point and each azimuthal wavenumber:
\begin{equation}\label{eqn:advanced-etd-linear-op-full}
\begin{aligned}
    \mathsfbi{L}(r, m) &=
    \begin{bmatrix}
        \mathsfbi{L}_H & \bm{0} \\
        \bm{0} & \mathsfbi{L}_V
    \end{bmatrix} \\
    &=
    \begin{bmatrix}
        -\iu m \bar{\Omega}(r) & 2\Omega + 2\bar{\Omega}(r) & 0 & 0 \\
        -2\Omega - \bar{\zeta}(r) & -\iu m \bar{\Omega}(r) & 0 & 0 \\
        0 & 0 & -\iu m \bar{\Omega}(r) & -1 \\
        0 & 0 & \bar{N}^2 & -\iu m \bar{\Omega}(r)
    \end{bmatrix}.
\end{aligned}
\end{equation}
We can therefore diagonalize and exponentiate this system at each $(r,m)$-pair independently without coupling the global domain.

Because the horizontal ($\mathsfbi{L}_H$) and vertical ($\mathsfbi{L}_V$) blocks cleanly decouple, they are diagonalized separately. For the horizontal block, the eigenvalues mathematically capture the Doppler-shifted epicyclic frequencies, yielding the following diagonal and eigenvector matrices:
\begin{subequations}\label{eqn:etd-h}
\begin{align}
    \mathsfbi{J}_H &= \begin{bmatrix}
        -\iu(m\bar{\Omega} + \sqrt{\Delta}) & 0 \\
        0 & -\iu(m\bar{\Omega} - \sqrt{\Delta})
    \end{bmatrix}, \label{eqn:etd-jh} \\
    \mathsfbi{S}_H &= \begin{bmatrix}
        \frac{\iu\sqrt{\Delta}}{2\Omega+\bar{\zeta}} & -\frac{\iu\sqrt{\Delta}}{2\Omega+\bar{\zeta}} \\
        1 & 1
    \end{bmatrix}, \quad
    \mathsfbi{S}_H^{-1} = \begin{bmatrix}
        -\iu\frac{2\Omega+\bar{\zeta}}{2\sqrt{\Delta}} & \frac{1}{2} \\
        \iu\frac{2\Omega+\bar{\zeta}}{2\sqrt{\Delta}} & \frac{1}{2}
    \end{bmatrix},
\end{align}
\end{subequations}
where the generalized Rayleigh criterion $\Delta(r)$ is defined as
\begin{equation}\label{eqn:rayleigh-discriminant}
    \Delta(r) \equiv 2(\bar{\zeta}(r)+2\Omega)(\bar{\Omega}(r)+\Omega).
\end{equation}
For the vertical block, the eigenvalues correspond to the Doppler-shifted buoyancy frequencies, evaluating to:
\begin{subequations}\label{eqn:etd-v}
\begin{align}
    \mathsfbi{J}_V &= \begin{bmatrix}
        -\iu(m\bar{\Omega} + \bar{N}) & 0 \\
        0 & -\iu(m\bar{\Omega} - \bar{N})
    \end{bmatrix}, \label{eqn:etd-jv}\\
    \mathsfbi{S}_V &= \begin{bmatrix} -\frac{\iu}{\bar{N}} & \frac{\iu}{\bar{N}} \\ 1 & 1 \end{bmatrix}, \quad
    \mathsfbi{S}_V^{-1} = \begin{bmatrix} \frac{\iu \bar{N}}{2} & \frac{1}{2} \\  -\frac{\iu \bar{N}}{2} & \frac{1}{2} \end{bmatrix}.
\end{align}
\end{subequations}
Notably, these vertical eigenvectors remain identical to those of the basic unsheared ETD scheme, as the advective cross term physically applies a Doppler shift to the phase accumulation rate without altering the fundamental vertical eigenstructure.

This complete analytical diagonalization reveals a profound physical elegance behind the ETD framework: the fundamental stability limits and resonance characteristics of the fluid are intrinsically encoded within the mathematical structure of the integration operators. The matrix $\mathsfbi{J}_H$ yields purely imaginary (oscillatory) eigenvalues if and only if $\Delta(r) > 0$ universally. This guarantees that the numerical scheme rigorously respects the physical requirement for centrifugal stability of the background flow, ensuring the analytical eigenvectors remain well-defined and avoiding the mathematical singularity at $\Delta(r) = 0$. Similarly, the vertical formulation becomes singular when $\bar{N} = 0$; however, this unstratified limit falls outside the scope of the stratified regimes considered in the present study. 

Furthermore, the formulation naturally accommodates critical layer phenomena. As evidenced by (\ref{eqn:etd-jh}) and (\ref{eqn:etd-jv}), exact zero eigenvalues will occur at critical radii, $r_c$, where the local Doppler-shifted frequencies vanish: $\bar{\Omega}(r_c) = \pm \bar{N}/m$ (baroclinic critical layers) or $\bar{\Omega}(r_c) = \pm \sqrt{\Delta(r_c)}/m$ (barotropic critical layers). Physically, these critical layers correspond to regions of local wave resonance, absorption, and momentum transfer \citep{booker_bretherton_1967, maslowe_1986}, and they have been shown to incite instability dynamics such as the zombie vortex instability \citep{Marcus_2013,Marcus_2015,Barranco_2018,Balmforth_2021,Wohlever_2025}. Within our ETD framework, these dynamically vital regions naturally emerge as exact zero eigenvalues in the linear operator. Because the $\mathsfbi{NL}$ operator handles these resonant limits via the analytical L'Hôpital limit defined previously in (\ref{eqn:etd-eigval-limit}), the numerical scheme inherently captures the critical layer dynamics without requiring any \textit{ad hoc} numerical intervention.

Operationally, this advanced ETD scheme follows the exact same algorithmic recipe detailed in \S\ref{sec:basic-etd}, with three key implementation distinctions. First, because the linear cross terms are now absorbed analytically into $\mathsfbi{L}(r,m)$, the explicitly evaluated forcing vector is strictly reduced to the nonlinear self-advection: $\bm{f}^{t_N} = \mathcal{N}(\bm{v}^{t_N})$. Second, because the matrices $\mathsfbi{E}$ and $\mathsfbi{NL}$ are radially dependent, the ETD advancement step must be computed in the hybrid PFF space, applying the explicit updates at each radial collocation point before transforming back to the fully spectral FFF space for the diffusion solve and pressure extraction. Third, by fully diagonalizing the advective cross terms, the instantaneous pressure balance of the $z$-invariant ($k=0$) modes is organically assimilated into the governing eigenvalue structure, which entirely eliminates the need for the manual non-rotating overrides required by the basic scheme.

Most importantly, this advanced ETD scheme achieves the removal of the background flow advective stability constraint. Because both the restorative wave forces and the background shear interactions are integrated analytically, the allowable explicit time step is no longer restricted by the rapid frequencies of the background flow nor the background rotation or stratification rate. Instead, the explicit stability limit is dictated by the temporal truncation error of the fully nonlinear perturbation advection. This paradigm shift allows the integration time step to be scaled according to the slow, macroscopic evolution of the physical instabilities of interest, rather than the fast background kinematics, rendering long-term, high-resolution integration computationally highly efficient.

\subsection{Numerical stabilization, aliasing, and outgoing waves}\label{sec:stabilization}
For pseudo-spectral methods, evaluating nonlinear advective products in the physical collocation space intrinsically generates high-frequency harmonics. In high-Reynolds-number or inviscid simulations, the physical diffusion operator ($\mathcal{L}_{\mathrm{diff}}$) is typically insufficient to dissipate this forward energy cascade. To prevent artificial energy pile-up at the truncation limits and the subsequent back-scattering of aliasing errors, the numerical framework provides two distinct stabilization paradigms.

For the strictly periodic azimuthal and axial directions, which are discretized using uniform Fourier series, one can explicitly apply the standard $2/3$ de-aliasing rule to geometrically truncate the unresolved high-frequency harmonics. Alternatively, and specifically mandated for the semi-infinite radial direction, where the non-uniform Gauss-Legendre quadrature grid renders a strict geometric truncation less effective, one can retain the full spectral capacity and rely entirely on a dynamically adjusted, scale-selective hyperviscosity and hyperdiffusivity filter to exponentially suppress the forward cascade before it reaches the grid scale \citep{canuto_2007,boyd_1998,barranco_2006}.

When employing the hyperviscosity operator, rather than applying computationally rigid, high-order differential operators (e.g., $\nabla^{p}$), we implement it as an exact integrating factor applied to the spectral modal amplitudes at the conclusion of each time step. For a given scalar variable $\phi$ (representing a component of the state vector), the generalized spectral filter takes the form:
\begin{equation}
    \phi^{t_{N+1}}_{\mathrm{filtered}} = \phi^{t_{N+1}} \exp\left( - \Delta t \sum_{j \in \{r, \theta, z\}} \nu_{\mathrm{hyp},j} \mathcal{M}_j(w_{\mathrm{op},j}) \left(\frac{w_{\mathrm{op},j}}{w_{\mathrm{op},j}^{\max}}\right)^{p/2} \right),
    \label{eqn:hyperviscosity}
\end{equation}
where $\nu_{\mathrm{hyp},j}$ are the directional dissipation coefficients (hyperviscosity and hyperdiffusivity), and $w_{\mathrm{op},j}$ represents the specific directional eigenvalue of the underlying spectral basis operators (i.e., $w_{\mathrm{op},\theta} = m^2$, $w_{\mathrm{op},z} = k^2$, and $w_{\mathrm{op},r} = n(n+1)$, where $n$ is the degree of the associated Legendre polynomial). The masking function, $\mathcal{M}_j(w_{\mathrm{op},j})$, is a smooth hyperbolic tangent step profile designed to strictly evaluate to zero for the lower wavenumbers, ensuring the large-scale inertial physics remain completely undamped.

To ensure the dissipation rate perfectly matches the dynamically varying nonlinear cascade without over-damping the flow, the coefficients $\nu_{\mathrm{hyp},j}$ can be adjusted dynamically every hundreds of time steps via a Proportional-Derivative (PD) feedback control loop. At diagnostic intervals, we leverage the Parseval's theorem to evaluate the 1D kinetic and available potential energy spectra along the periodic azimuthal and axial directions (as detailed in Appendix~\ref{app:energy}). We extract the spectral shape of the high-wavenumber tail by projecting the logarithmic energy spectra, $\log_{10} E(x)$, onto the first and second Legendre polynomials ($P_1$ and $P_2$), where $x \in [-1, 1]$ maps across the tail wavenumbers. These projections quantify the local spectral slope and curvature, respectively. A positive curvature indicates an upward concavity (an energy bottleneck), while the slope indicates the bulk dissipation rate. The dissipation coefficients are then actively updated based on the deviation from a healthy, theoretical inertial-range target state:
\begin{equation}
    \nu_{\mathrm{hyp},j}^{t+\Delta t} = \nu_{\mathrm{hyp},j}^{t} \left[ 1 + K_p \left(C^t - C^{t-\Delta t}\right) + K_d \left(C^t - C_{\mathrm{target}}\right) \right],
\end{equation}
where $C$ is the measured spectral metric (curvature or slope), and $K_p$ and $K_d$ are the proportional and derivative gains. For the azimuthal and axial directions, the corresponding metrics and damping coefficients can be directly evaluated from their respective 1D spectra. Because the non-periodic mapped radial basis precludes a direct 1D Parseval spectral density, the radial metric $C_r$ is formulated as a weighted average of the azimuthal and axial metrics, biased strictly towards the direction exhibiting the more severe spectral pile-up. This control architecture dynamically ramps up the hyperviscosity and hyperdiffusivity exactly when and where spectral blocking is detected, and smoothly relaxes them once the turbulent cascade is properly resolved.

Finally, we must manage outgoing waves radiating toward the domain boundaries. For the unbounded radial direction, the mapped Legendre basis natively provides an algebraic decay as $r \to \infty$. We deliberately avoid implementing a radial sponge layer since applying a spatially varying damping profile in the radial direction of a strongly sheared background flow artificially injects localized spurious vorticity, which can trigger non-physical instabilities. 

On the other hand, to safely absorb waves reaching the artificial periodic boundaries in the axial direction, we implement a physical 1D axial sponge layer, adapted from \citet{Mahdinia_2017}. A damping term $-D(z)\phi$ is appended to the evolution equations, utilizing the spatial profile:
\begin{equation}
    D(z) = \frac{1 - T(z)}{\tau},
\end{equation}
where $T(z)$ is a smooth hyperbolic tangent top-hat function maintaining $T=1$ (zero damping) throughout the interior domain, and $\tau$ dictates the damping timescale near the axial boundaries. To preserve the global second-order temporal accuracy of the ETD scheme without violating incompressibility, this damping is integrated in the hybrid PFP space at the end of each fractional step:
\begin{equation}
    \phi^{t_{N+1}}_{\text{sponged}} = \phi^{t_{N+1}} \exp\left(-D(z)\Delta t\right).
\end{equation}
Most importantly, this exponential scaling is applied exclusively to the scalar potentials ($\psi, \chi$) and the buoyancy perturbation ($b'$), which guarantees that the reconstructed 3D velocity field remains strictly divergence-free.

\section{Numerical tests}\label{sec:numerical-tests}

\subsection{Parallel performance and scaling}\label{sec:parallel-performance}

The present numerical framework is developed as an expansion to the open-source MLEGS library, which was originally designed for solving the incompressible Navier-Stokes equations. Tailored specifically for the Boussinesq equations under background shear and stratification, this ETD module (to be released in a forthcoming version update) introduces cross-platform compatibility (supporting Linux and macOS environments) for deployment ranging from local workstations to large-scale computing clusters. To isolate algorithmic scaling properties from the hardware-induced latency variances typical of multi-node network interconnects, the performance profiling here is strictly confined to a single high-performance computing node on the Purdue Anvil supercomputer \citep{ANVIL}. The compute node is equipped with dual-socket AMD EPYC 7763 processors clocked at 2.45GHz, providing a total of 128 physical cores with a layered cache architecture including $32$KB of L1 data (L1d) and instruction (L1i) caches, $512$KB of L2, and $32$MB of L3 cache.

The domain is partitioned via a 2D pencil decomposition strategy using the Message Passing Interface (MPI) . To compute different physical operators, the algorithm transitions through multiple data spaces. In the physical (PPP) and intermediate (PFP) spaces, the radial and axial dimensions are distributed across the processor grid, while the azimuthal dimension remains complete (local) on each processor. Conversely, the PFF and FFF spaces distribute the azimuthal and axial dimensions, rendering the radial dimension entirely complete. Consequently, the explicit nonlinear pointwise evaluations ($\mathcal{N}$) are computed locally in the PPP space, and the ETD matrix advancements ($\mathsfbi{E}, \mathsfbi{NL}$) are evaluated in the PFF space without requiring any inter-process communication. The poloidal-toroidal projection ($\mathbb{P}$), which is functionally identical to a pressure solver in typical numerical codes to enforce incompressibility, is similarly evaluated locally in the spectral space. The primary parallel communication occurs during the space conversions, necessitating global MPI all-to-all swaps between the PFP and PFF spaces to transpose the data layout.

The strong scaling simulations are executed using a physical collocation grid of $256 \times 256 \times 128$. Because the azimuthal transform utilizes a real-to-complex FFT, the 256 physical azimuthal points are stored as $128$ complex Fourier modes. Furthermore, to reserve spectral capacity for high-order radial differentiation, the effective spectral resolution is internally truncated to $250 \times 128 \times 127$ modes. This problem size provides a substantial computational workload to evaluate memory bandwidth saturation while fitting comfortably within the shared system memory.

\begin{figure}
    \centering
    \includegraphics[height=0.3\textheight]{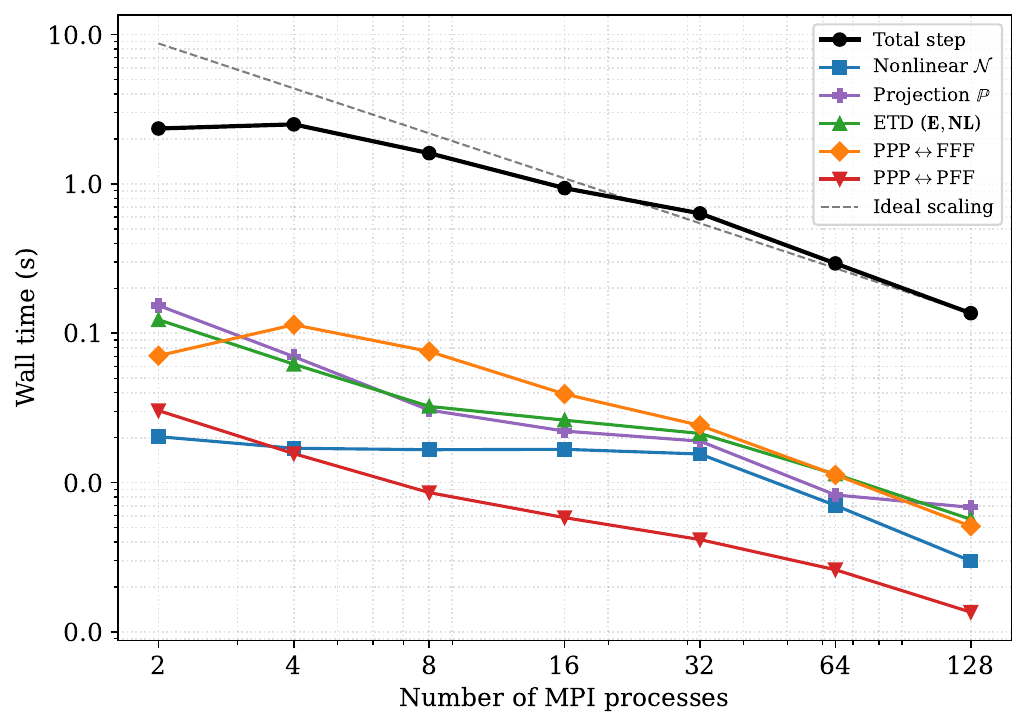}
    \caption{Single-node strong scaling performance for a $256 \times 256 \times 128$ physical grid on dual-socket AMD EPYC 7763 processors ($128$ total cores). Legend labels denote the total step time, the pointwise nonlinear evaluation ($\mathcal{N}$), the ETD advancement ($\mathsfbi{E}, \mathsfbi{NL}$), the full space conversions (PPP $\leftrightarrow$ FFF), the localized radial expansions, and the solenoidal projection ($\mathbb{P}$). Note that the plotted timings for the space conversions and radial expansions represent the average duration per single operation, whereas a complete integration step encompasses multiple such transformations. The dashed line represents ideal linear scaling.}
    \label{fig:scaling}
\end{figure}

The actual hardware-software interaction is revealed in the strong scaling results depicted in figure~\ref{fig:scaling}. For the explicit nonlinear computation ($\mathcal{N}$), one would theoretically expect ideal linear scaling since the pointwise multiplications require zero MPI communication. However, because its computation traverses the entire physical grid and involves heavy, continuous memory access, it rapidly saturates the memory controllers. This is reflected in the scaling curve, where the nonlinear evaluation speedup stalls severely between $2$ and $32$ processors. Performance only recovers into an efficient scaling regime at $64$ to $128$ cores, at which point the partitioned local sub-arrays become sufficiently small to reside entirely within the CPU caches.

In contrast, the ETD advancement ($\mathsfbi{E}, \mathsfbi{NL}$) and projection ($\mathbb{P}$) operators involve denser algebraic operations on smaller localized blocks. Consequently, they maintain near-ideal scaling from $2$ to $8$ cores before eventually hitting a delayed efficiency plateau. This mid-range degradation is attributed to a combination of memory bandwidth saturation and minor computational load imbalances, arising from the indivisible number of spectral modes (e.g., the odd truncation limit of $127$ axial modes) distributed unevenly across the MPI processor grid. Both operators similarly recover at higher core counts. The localized radial transforms (i.e., radial spectral expansion and reverse) maintain consistent scaling throughout. Meanwhile, the full space conversion (PPP $\leftrightarrow$ FFF), which dictates the primary MPI communication cost due to the algorithm's pseudo-spectral nature, experiences an initial latency penalty from $2$ to $4$ cores, likely due to crossing internal NUMA or socket boundaries, before scaling efficiently at larger process counts. These scaling results demonstrate that, during the active time-stepping phase, the linear matrix advancements from ETD introduce no significant performance penalty. Because these operators act via localized block-diagonal multiplications in the spectral space, they scale highly efficiently and avoid the severe communication overhead associated with global data transposes. 

To provide a complete assessment of the ETD framework's computational footprint, it is also necessary to mention its initialization overhead. While the scheme requires pre-computing the integration operators during simulation startup, this associated cost is mathematically negligible. The full ETD framework requires diagonalization for each radial collocation point and azimuthal wavenumber; however, utilizing the analytical symbolic expressions (\ref{eqn:etd-h}) and (\ref{eqn:etd-v}) ensures the computational complexity of this initialization phase remains strictly $\mathcal{O}(N_r \cdot M)$, where $N_r$ and $M$ denote the number of radial collocation points and azimuthal wavenumbers, respectively.

Ultimately, this combination of limited initialization overhead and highly efficient localized time-stepping underscores the computational advantage of the ETD framework over traditional integrators. Given that the pointwise nonlinear calculations hit a severe memory bottleneck even at lower core counts, integrating conventional mixed IMEX schemes, such as the AB2-CN, would require the allocation, storage, and repeated memory retrieval of multiple historical right-hand-side arrays. This additional memory traffic would heavily exacerbate the observed scaling degradation. By strictly relying on the current state vector via localized algebraic updates, the ETD scheme reduces this excess memory latency, proving computationally superior in memory-bound regimes. Furthermore, the robust recovery and near-ideal scaling achieved at higher core counts indicate that the 2D pencil decomposition parallelization strategy is highly successful. By effectively distributing the domain such that the local sub-arrays fit entirely within the processors' high-speed cache architecture, the numerical framework demonstrates excellent parallel efficiency, making it well-optimized for actual production deployment on large-scale computing clusters.


\subsection{Lamb-Oseen vortex in a stratified fluid}\label{sec:lamb-oseen-validation}
To validate the numerical framework and evaluate the performance of the exponential time differencing (ETD) scheme, we use a stably stratified Lamb-Oseen vortex as the background shear flow. Without loss of generality, we consider a generic vortex with core radius $a = 1$ and circulation $\Gamma=2\pi$, such that its azimuthal velocity profile is given by:
\begin{equation}
    \bar{U}_\theta(r) = \frac{1-e^{-r^2}}{r}.
\end{equation}
For the numerical experiments detailed below, the Brunt-Väisälä frequency ranges from $\bar{N} = 0.1$ to $5.0$. We further consider the vortex in a non-rotating environment ($\Omega = 0$). While the inclusion of a background rotation will fundamentally alter the stability landscape by modifying the Coriolis restoring forces, shifting the generalized Rayleigh discriminant $\Delta(r)$, and potentially introducing rotation-dependent mechanisms such as the stratorotational instability, setting $\Omega = 0$ serves two specific validation purposes. First, it isolates the Baroclinic shear-stratification interaction, aligning our numerical setup precisely with the documented benchmark of \citet{dizes_2008} for a clean physical cross-reference. Second, from an algorithm perspective, the background rotation enters the linearized governing equations exclusively as an additive constant to the background axial vorticity within the horizontal transformation matrix ($\mathsfbi{S}_H$ and $\mathsfbi{S}_H^{-1}$). Therefore, setting a zero background rotation does not simplify the algebraic structure, the cross-variable coupling, or the numerical stiffness of the linear operators. The full functional machinery of the ETD diagonalization is therefore fully exercised and validates even in the absence of system rotation.

\subsubsection{Linear stability and spatial validation}\label{sec:spatial-validation}
We first validate the linear spatial operators ($\mathcal{L}_{\mathrm{wave}} + \mathcal{L}_{\mathrm{cross}}$) and the global accuracy of the mapped Legendre discretization by computing the discrete global eigenmodes of the stratified Lamb-Oseen vortex. We emphasize that while the nonlinear operator $\bm{u}'\times\bm{\omega}'$ has been thoroughly validated in preceding studies utilizing the mapped associated Legendre basis \citep{matsushima_1997}, the linear operators are the primary focus here, as they constitute the functional foundation of the ETD scheme.

Assuming normal mode disturbances of the form $\bm{v}'(\bm{x}, t) = \tilde{\bm{v}}(r) \exp[\iu(m\theta + kz) + \sigma t]$, the linearized, inviscid Boussinesq equations reduce to the generalized eigenvalue problem:
\begin{equation}
    \sigma \tilde{\bm{v}} = \mathsfbi{L}(r, m) \tilde{\bm{v}} - \tilde{\bm{\Pi}}.
    \label{eqn:evp-formulation}
\end{equation}
Here, $\sigma \equiv \lambda + \iu\omega$ is the complex eigenvalue, where $\lambda$ denotes the growth rate and $\omega$ is the wave frequency. The velocity eigenvector, $\tilde{\bm{u}} = [\tilde{u}_r, \tilde{u}_\theta, \tilde{u}_z]^\intercal$, is subject to the linear solenoidal constraint:
\begin{equation}
    \frac{1}{r}\frac{\mathrm{d}}{\mathrm{d}r}(r\tilde{u}_r) + \frac{\iu m}{r}\tilde{u}_\theta + \iu k \tilde{u}_z = 0.
\end{equation}
To enforce this divergence-free condition and simultaneously eliminate the pressure perturbation $\tilde{\bm{\Pi}}$, the global eigenvalue matrix is constructed using the poloidal-toroidal formulation. Specifically, we define spectral test functions in the reduced scalar basis $(\tilde{\psi}, \tilde{\chi}, \tilde{b})$ and expand these potentials in the mapped Legendre space to reconstruct the corresponding primitive variables. The linear operator is then applied in the PFF space at each radial collocation point. Finally, the resulting state is projected back to the poloidal-toroidal form in the FFF space, yielding the corresponding columns of the global eigenvalue matrix.

At each collocation point, the local linear dynamics are governed by the purely algebraic operator $\mathsfbi{L}(r, m)$. Exploiting the analytical diagonalization derived for our advanced ETD scheme in \S\ref{sec:advanced-etd}, this operator is evaluated via the similarity transformation:
\begin{equation}
    \mathsfbi{L}(r, m) = \mathsfbi{S}(r) \mathsfbi{J}(r, m) \mathsfbi{S}^{-1}(r).
\end{equation}
The diagonal matrix $\mathsfbi{J}(r, m)$ explicitly separates into a restorative component, which remains constant across all spatial wavenumbers $(m, k)$, and an advective Doppler shift that scales solely with the azimuthal wavenumber $m$:
\begin{equation}
    \mathsfbi{J}(r, m) = \mathrm{diag}\left( \omega^\pm_H(r),\, \pm \iu\bar{N}\right) - \iu m\bar{\Omega}(r)\mathsfbi{I},
\end{equation}
where the horizontal epicyclic frequencies are defined as $\omega^\pm_H(r) = \pm\iu\sqrt{\Delta(r)}$ with $\Delta(r)$ being the generalized Rayleigh discriminant given by (\ref{eqn:rayleigh-discriminant}), and $\mathsfbi{I}$ is the $4\times 4$ identity matrix. These epicyclic frequencies dictate the natural, inertial restoring oscillations of a fluid parcel displaced radially within the rotating background shear, and are governed entirely by the generalized Rayleigh discriminant. Because the constituent matrix $\mathsfbi{L}(r, m)$ is completely independent of the axial wavenumber $k$, our global eigenvalue solver can efficiently scan across a large parameter space without the need to re-calculate the linear operator every single time.

Before discussing the eigenvalue results, it is highly instructive to analyze the system using a WKBJ framework \citep{dizes_2008}, which helps classify the resulting eigenmodes and understand their spatial structures. We define the Doppler-shifted complex eigenvalue as $\Phi(r) \equiv \sigma + \iu m\bar{\Omega}(r)$. In the large-$k$ asymptotic limit, the spatial envelope of a wave is governed by the local radial wavenumber $\beta(r)$, given by
\begin{equation}
    \beta(r) = k\sqrt{\frac{\Phi^2(r) + \Delta(r)}{\Phi^2(r) + \bar{N}^2}}.
\end{equation}
The lowest-order WKBJ approximation indicates that a mode is spatially oscillatory where $\beta^2 > 0$ and evanescent where $\beta^2 < 0$. The spatial boundaries separating these regions are the kinematic turning points $r_t$, defined by the roots of the numerator: $\Phi^2(r_t) + \Delta(r_t) = 0$, or equivalently $\Phi(r_t) = \omega^\pm_H(r_t)$, matching the local epicyclic frequencies encoded in our ETD operator.

Furthermore, the continuous equations possess two critical layer singularities (see Appendix~\ref{app:evp} for details). The baroclinic critical layer occurs where $\Phi^2(r_c) = \bar{N}^2$, indicating a resonance with the background stratification ($\beta \to \infty$). The barotropic critical layer occurs where the wave is in perfect co-rotation with the local background flow:
\begin{equation}
    \Phi(r_c) = \omega + m\bar{\Omega}(r_c) = 0.
\end{equation}
Comparing the above equation with the definition of the turning points reveals that the frequency curve of the barotropic critical layer will always sit between the two epicyclic frequencies ($\omega^\pm_H$). As studied in \citet{dizes_2008}, for a typical ring mode of the Lamb-Oseen vortex under strong stratification, there are generally three turning points, and the barotropic critical layer is located exactly inside the evanescent (negative $\beta^2$) region bounded between two of these turning points. This implies that the turning points act as potential barriers that exponentially shield the active wave from the singular critical layer.

\begin{figure}
    \centering
    \includegraphics[width=.95\textwidth]{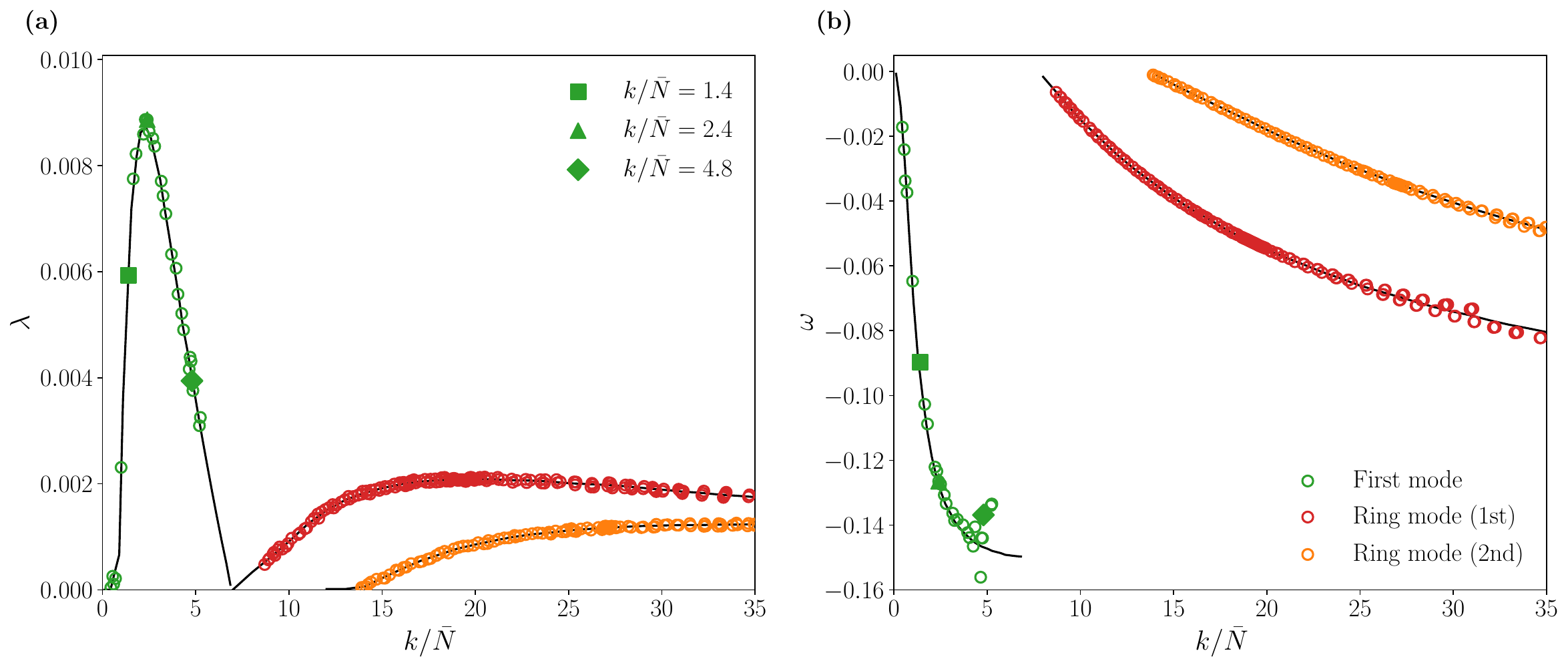}
    \caption{Discrete unstable modes for a stratified Lamb-Oseen vortex at $\bar{N}=5$. (a) Growth rates $\lambda$ and (b) wave frequencies $\omega$ as functions of the scaled axial wavenumber $k/\bar{N}$. Open circles represent the eigenvalues computed by the present global spectral eigenvalue solver. Three points on the first mode branch ($k/\bar{N} = 1.4, 2.4, 4.8$) are distinctly marked for further structural analysis. Solid black lines denote the benchmark results of \citet{dizes_2008} obtained via a shooting method with complex contour deformation. The numerical spatial discretization employs $N_r = 252$ radial collocation points, a radial spectral truncation of $N = 244$, and a mapping parameter $L = 6.0$. Only the first three unstable branches are plotted.}
    \label{fig:dispersion_validation}
\end{figure}

We benchmark our eigenvalue solver against the numerical results of \citet{dizes_2008}, who utilized a shooting method with complex contour integration to obtain the dispersion curves of the helical ($m=1$) modes at $\bar{N}=5$. As shown in figure~\ref{fig:dispersion_validation}, our global spectral eigenvalue solver accurately captures the three unstable branches. An error of less than $1\%$ is achieved for all wavenumbers of the second ring mode, and for wavenumbers up to $k/\bar{N} = 25$ for the first ring mode. At higher wavenumbers, the wave frequencies of the first ring mode obtained by our real-axis solver begin to show minor spectral oscillations (with errors remaining below $2.5\%$). These minor deviations are a direct consequence of evaluating the linear critical layer singularity on the purely real collocation axis. As analyzed in detail below, this real-axis evaluation also governs the spectral behavior of the primary first mode, which undergoes a unique structural evolution that fully exposes this mathematical singularity.

\begin{figure}
    \centering
    \includegraphics[width=\textwidth]{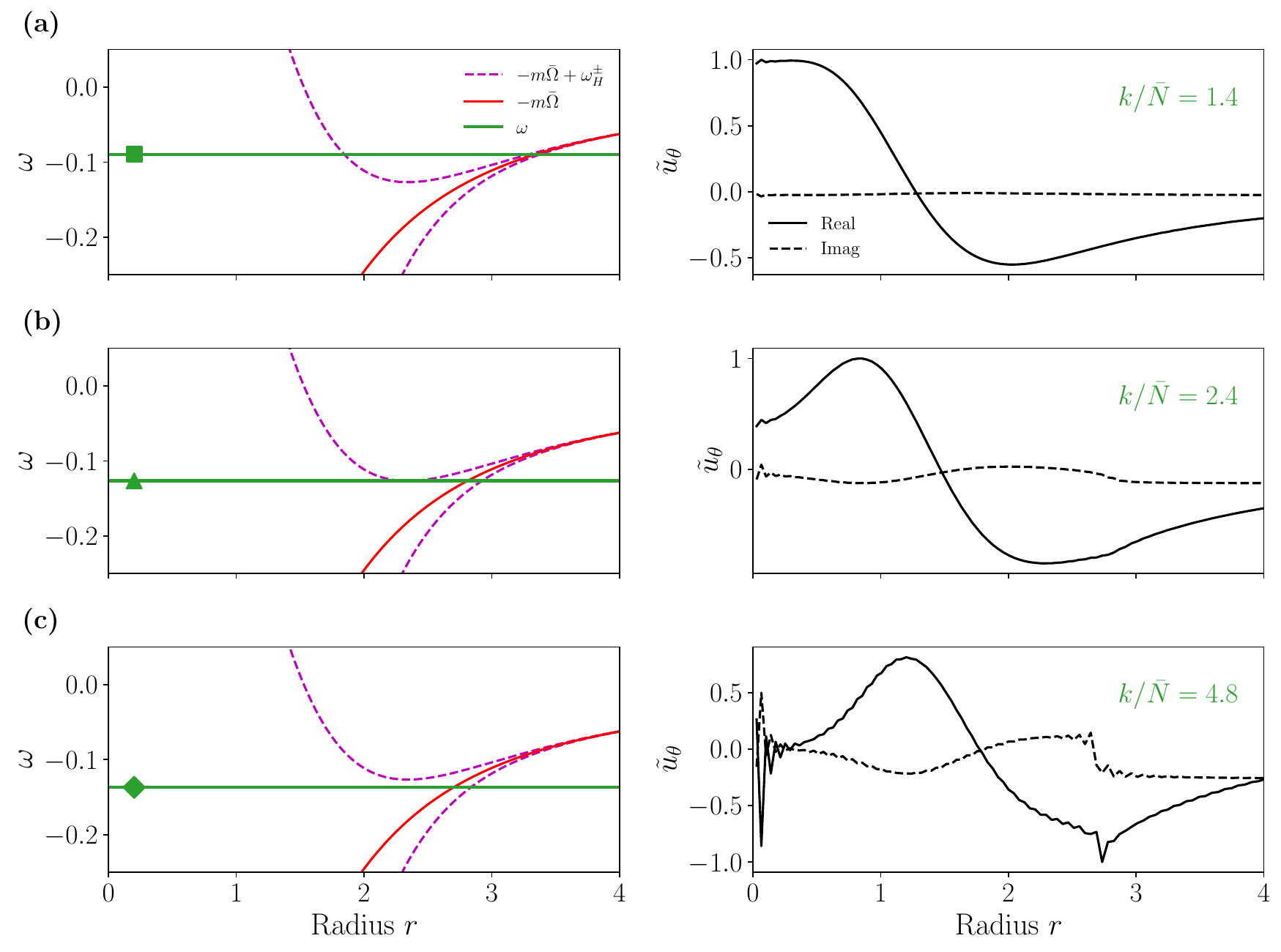}
    \caption{Frequency maps (left column) and normalized spatial eigenmode profiles $\tilde{u}_\theta$ (right column) tracking the topological evolution of the first mode at the three distinct scaled wavenumbers, as marked in figure~\ref{fig:dispersion_validation}: (a) $k/\bar{N} = 1.4$, (b) $k/\bar{N} = 2.4$, and (c) $k/\bar{N} = 4.8$. The left panels plot the radial profiles of the local epicyclic frequencies (dashed magenta lines) and the co-rotation frequency (solid red line), alongside the mode's constant wave frequency $\omega$ (horizontal green line). The intersections of the horizontal wave frequency line with these respective curves define the radial locations of the turning points and the barotropic critical layer. At $k/\bar{N} = 4.8$, the loss of the inner turning points exposes the mode directly to the barotropic critical layer, triggering the Gibbs-like spectral ringing observed in the spatial profile.}
    \label{fig:eigenmode_profiles}
\end{figure}

As pointed out by \citet{dizes_2008}, the first mode (the most unstable branch) exhibits unique physical characteristics that separate it from the standard ring mode family. Figure~\ref{fig:eigenmode_profiles} tracks the structural evolution of this mode across three distinct stages: before its peak growth rate ($k/\bar{N} = 1.4$), at its peak ($k/\bar{N} = 2.4$), and after it visibly detaches from the theoretical curve ($k/\bar{N} = 4.8$).

Initially (figure~\ref{fig:eigenmode_profiles}a), the mode behaves similarly to a standard ring mode; its wave frequency intersects the $\omega^-_H$ curve at two distinct inner turning points and the $\omega^+_H$ curve at an outer turning point. Because the barotropic critical layer ($\Phi=0$) resides between the $\omega^+_H$ and $\omega^-_H$ curves, it is located within the resulting evanescent region ($\beta^2 < 0$). Correspondingly, its mode profiles are smooth and well-resolved. In contrast, as the wavenumber increases and the mode reaches its peak growth rate (figure~\ref{fig:eigenmode_profiles}b), its wave frequency $\omega$ drops to the local minimum (valley) of the $\omega^-_H$ curve. At this juncture, the two inner turning points collide. Eventually, as the frequency drops further (figure~\ref{fig:eigenmode_profiles}c), these inner turning points completely vanish. Without this inner evanescent potential barrier, the oscillatory region extends entirely to the center of the vortex ($r=0$), transforming the disturbance into a core mode. More critically, the removal of the turning points strips away the spatial shielding. The barotropic critical layer ($\Phi=0$) is no longer hidden in the evanescent zone; it emerges directly inside the active oscillatory region as a naked mathematical singularity.

While \citet{dizes_2008,dizes_2010} bypass the critical layer singularity using complex contours or mappings, and \citet{lee_2023} optimizes the spatial mapping parameter $L$ to smoothly resolve it, our spectral discretization is evaluated strictly on the real radial axis to serve the physical nonlinear ETD scheme. Attempting to fit the resulting singular, logarithmic phase jump associated with this unshielded critical layer using globally smooth polynomials naturally triggers the Gibbs-like spectral ringing observed in the $\tilde{u}_\theta$ profiles in figure~\ref{fig:eigenmode_profiles}c. This ringing directly accounts for the deviations from the numerical eigenvalues of \citet{dizes_2008} at high wavenumbers. However, while critical layers are physical resonance phenomena, the mathematical singularity introduced at these locations exists only in the linear, inviscid limit. Because finite-amplitude nonlinear effects and viscosity inherently smoothen and regularize such resonances in full temporal simulations, complex mapping is physically unwarranted. As our eigenvalue solver is constructed strictly to validate our real-axis spatial operators, the observed spectral behaviors are entirely expected, and the real-axis resolution remains highly satisfactory for the target nonlinear physics.

\subsubsection{Temporal convergence and energy conservation}\label{sec:numerical-validation}
To confirm the temporal accuracy of the ETD scheme, we conduct a fractional error convergence test. The combination of AB2 for nonlinear forcing extrapolation and CN for diffusion approximation, as given by (\ref{eqn:etd-approx-intermediate}), suggests a global temporal error of $\mathcal{O}(\Delta t^2)$ for the overall ETD scheme. 

\begin{figure}
    \centering
    \includegraphics[width=.5\textwidth]{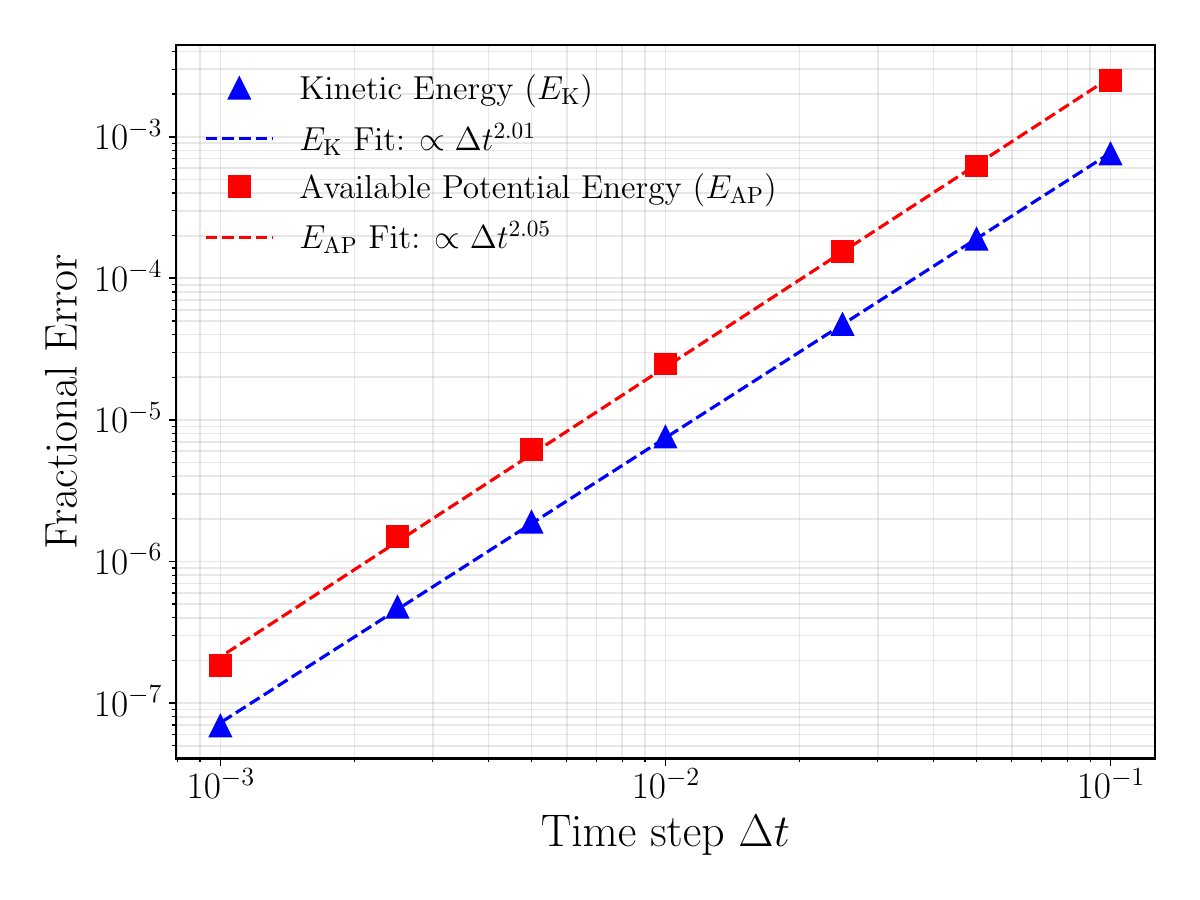}
    \caption{Temporal convergence of the ETD scheme. The fractional errors in global kinetic energy ($E_\mathrm{K}$) and available potential energy ($E_\mathrm{AP}$) follow the theoretical second-order scaling ($\propto \Delta t^2$) as $\Delta t$ is reduced. Errors are evaluated at $t=20$, corresponding to approximately $3.18$ core rotation periods ($T_{\text{core}} = 2\pi$) of the background Lamb-Oseen vortex, ensuring the measurement captures fully developed nonlinear dynamics.}
    \label{fig:convergence}
\end{figure}

To strictly isolate the temporal truncation error without contamination from spatial aliasing or boundary discontinuities, the convergence test is performed on a band-limited flow field. We discretize the domain using a spatial resolution of $(N_r, N_\theta, N_z) = (160, 256, 128)$ with a radial mapping parameter of $L=10.0$ and an axial domain length of $L_z = 10\pi$. To prevent nonlinear spectral blocking from polluting the temporal measurement, the spectral coefficients are de-aliased using a truncation of $(N, M, K) = (152, 84, 42)$, and a hyperviscosity of order $6$ is turned on to manage any energy pile-up at higher-order radial modes, as discussed in \S\ref{sec:stabilization}. 

The flow is initialized with a low-amplitude, 3D localized off-axis Gaussian perturbation superposed on the columnar background flow. While the perturbation's structure is derived using the cyclo-geostrophic and hydrostatic balance framework \citep{Mahdinia_2017}, we deliberately widen the axial length scale of the Gaussian envelope relative to a quasi-equilibrium. This artificial broadening ensures the perturbation is smoothly and fully resolved by the axial grid, circumventing the Gibbs ringing and spurious energy injections associated with under-resolved vertical gradients. Finally, a hyperviscosity filter is temporarily applied to the initial condition to scrub any residual spectral noise. 

We monitor the kinetic energy ($E_\mathrm{K}$) and the available potential energy ($E_\mathrm{AP}$) during each simulation, whose numerical evaluations are detailed in Appendix~\ref{app:energy}. As depicted in figure~\ref{fig:convergence}, after integrating for more than three vortex turn-around time, the fractional errors for $E_\mathrm{K}$ and $E_\mathrm{AP}$ decay with a pristine slope of $2$ on a log-log scale. This confirms that the exact matrix exponentiation of the linear operators preserves the global second-order accuracy without introducing stiff transients. Note that the absolute magnitude of the fractional error is higher for $E_\mathrm{AP}$ than $E_\mathrm{K}$; this is physically expected, as the explicit nonlinear advection of the strongly stratified buoyancy field ($\mathbf{u}' \cdot \nabla b'$) possesses a larger absolute truncation error pre-factor than the momentum advection.

\begin{figure}
    \centering
    \includegraphics[width=1.0\linewidth]{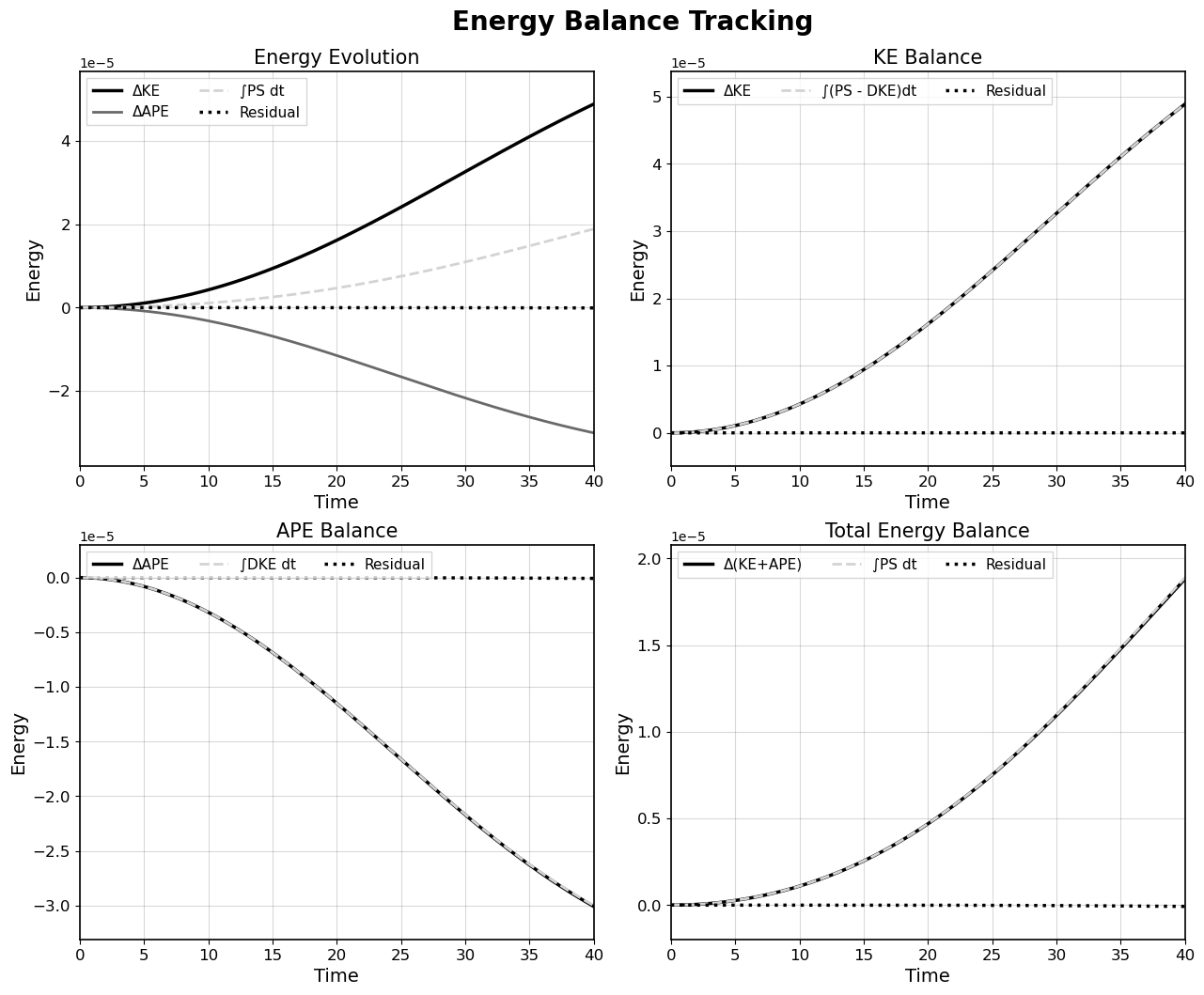}
    \caption{Energy balance tracking of the full ETD scheme during the nonlinear integration of the 3D Gaussian perturbation. To maintain inviscid stability and absorb the forward energy cascade, a sixth-order hyperviscosity filter is applied, with its amplitude dynamically adjusted based on the energy magnitude at the spectral tails. The figure comprises four panels: (top left) the temporal evolution of $E_\mathrm{K}$, $E_\mathrm{AP}$, and the cumulative shear production; (top right) the internal budget of $E_\mathrm{K}$, explicitly incorporating the buoyancy energy exchange; (bottom left) the internal budget of $E_\mathrm{AP}$, matching the buoyancy exchange; and (bottom right) the total mechanical energy balance. Note that the governing equations define the total energy evolution as $\dd_t(E_\mathrm{K} + E_\mathrm{AP}) = -\mathscr{E}_\mathrm{shear}$; here, we denote $\mathrm{PS} = -\mathscr{E}_\mathrm{shear}$.}
    \label{fig:ene-balance-full-etd}
\end{figure}

Beyond formal temporal accuracy, the nonlinear robustness of the ETD scheme is verified against the exact energy conservation laws given by (\ref{eqn:mechanical-energy-b'}). We track the energy balance of the Gaussian perturbation by monitoring the evolution of the kinetic and available potential energy relative to the cumulative shear production and the internal buoyancy energy exchange. These source and exchange terms are evaluated at each diagnostic step and subsequently integrated over time using the trapezoidal rule. 

As shown in figure~\ref{fig:ene-balance-full-etd}, the numerical scheme exhibits exceptional kinematic consistency. The bottom-right panel demonstrates that the net change in the total mechanical energy precisely matches the cumulative background shear production ($\mathrm{PS}$), strictly adhering to (\ref{eqn:conservation-law}). Furthermore, the individual energy budgets are properly maintained: the change in the disturbance kinetic energy balances the sum of the shear production and the buoyancy energy exchange (top right, consistent with (\ref{eqn:KE-balance})), while the change in the available potential energy mirrors the buoyancy exchange (bottom left, consistent with (\ref{eqn:APE-balance})). Ultimately, this strict adherence to the fundamental energy balances confirms that the fully diagonalized spectral ETD formulation successfully suppresses spurious numerical artifacts, enabling stable, physically faithful nonlinear integration at large time steps.

\section{Conclusions}\label{sec:conclusions}

This study presents a comprehensive pseudo-spectral numerical framework designed to simulate the three-dimensional Boussinesq equations for rotating, stably stratified flows within unbounded global cylindrical domains. To accurately capture the global geometry and isolate the fluid from artificial outer boundary reflections, the spatial discretization utilizes mapped associated Legendre polynomials in the radial direction coupled with Fourier series in the azimuthal and axial directions. This specific basis natively resolves the coordinate singularity at the origin and achieves spectral convergence for localized disturbances. Furthermore, by employing a poloidal-toroidal decomposition, the framework analytically enforces the divergence-free incompressibility constraint and eliminates the pressure variable from the governing momentum equations, streamlining the computational state space.

A primary contribution of this work is the formulation of a semi-analytical exponential time differencing (ETD) scheme to resolve the numerical stiffness inherent in rotating shear flows. Standard mixed implicit-explicit methods impose restrictive time step limits due to the high-frequency restorative forces of inertial-gravity waves and the rapid advective transport of the background azimuthal shear. By transitioning to the primitive velocity space, the proposed ETD scheme analytically diagonalizes the fully coupled linear operator, encompassing the background rotation, stratification, and the radially dependent advective cross terms. This exact integration steps over the fast linear timescales and embeds the physical resonance characteristics, including critical layers and centrifugal stability bounds, directly into the numerical operators.

The framework was validated through the simulation of a localized perturbation embedded within a stably stratified Lamb-Oseen vortex. The full ETD scheme effectively suppressed the high-frequency numerical artifacts that lead to instability in standard explicit treatments of background shear. By analytically integrating the linear cross terms, the stability constraint on the time step is decoupled from the fast background kinematics and is instead governed solely by the slow, macroscopic evolution of the nonlinear perturbation dynamics. The thermodynamic and kinematic consistency of the method was verified through continuous monitoring of the domain-integrated energy budgets, demonstrating adherence to the derived conservation laws for available potential energy, disturbance kinetic energy, and shear production.

Ultimately, this numerical methodology provides a mathematically consistent and computationally efficient foundation for investigating complex fluid phenomena that cannot be adequately captured by local Cartesian models. By rigorously enforcing global geometric constraints and circumventing advective stability limits, this framework is well-suited for long-term, high-resolution simulations of finite-amplitude hydrodynamic mechanisms, such as the zombie vortex instability, in astrophysical and geophysical environments.

\section{Acknowledgment}
We acknowledge the use of NSF ACCESS computational resources (ID: PHY220056).

\appendix

\section{Derivation of conservation laws}\label{app:conservation_derivation}
We consider a cylindrical domain ($0\leq r < \infty$) that is periodic in the azimuthal ($\theta$) and axial ($z$) directions. We enforce the physical boundary condition that all localized disturbance fields, such as the perturbation velocity $\bm{u}'$ and buoyancy $b'$, decay to zero as $r\rightarrow\infty$. While theoretical background shear profiles analogous to the local shearing box (e.g., constant vorticity profiles) may formally possess infinite total energy when integrated over an unbounded radial domain, rigorous conservation laws can be constructed by focusing on the bounded disturbance fields. For clarity, the following derivations omit viscous and diffusive terms, though the final energy budgets naturally extend to include these strictly dissipative sinks.

To establish the theoretical framework, we first consider a generalized, fully bounded velocity field $\bm{u}$ and buoyancy $b$ that decay at infinity. Taking the inner product of the inviscid momentum equation with $\bm{u}$ yields the evolution of the kinetic energy density. Noting that the Coriolis force does no work ($\bm{u}\cdot(2\Omega\dir{z}\cross\bm{u}) = 0$) and the nonlinear advection of kinetic energy can be expressed as a divergence, we obtain
\begin{equation}
\begin{aligned}
    \partial_t\left(\frac{1}{2}\bm{u}^2\right) &= -\bm{u}\cdot \grad \left(\frac{p}{\rho_0}+\frac{1}{2}\bm{u}^2\right) - b u_z \\
    &= -\div{\left[\left(\frac{p}{\rho_0}+\frac{1}{2}\bm{u}^2\right)\bm{u}\right]} - b u_z,
\end{aligned}
\end{equation}
where we have utilized the divergence-free condition $\div{\bm{u}} = 0$. Integrating over the domain volume $V$, the divergence term evaluates to zero by Gauss's theorem due to the periodic and radially decaying boundary conditions. 

For a linearly stratified fluid, the total buoyancy is $b = \bar{b}(z) + b'$, where the background profile is $\bar{b}(z) = -\bar{N}^2z + C$. The work done by this background stratification can be rewritten exactly as a divergence:
\begin{equation}
    \bar{b}u_z = (-\bar{N}^2z+C)u_z = \div{\left[\left(-\frac{1}{2}\bar{N}^2 z^2+Cz\right)\bm{u}\right]}.
\end{equation}
Because this term is a pure divergence, its volume integral also vanishes. This leaves the time rate of change of the generalized kinetic energy dependent solely on the buoyancy perturbation:
\begin{equation}
   \dv{E_\mathrm{K}}{t} = \partial_t \int_V\frac{1}{2}\bm{u}^2dV = -\int_V(b' u_z)\dd{V}.
\end{equation}
Similarly, the time rate of change of the available potential energy ($E_\mathrm{AP}$) associated with the buoyancy perturbation is calculated via the thermodynamic equation:
\begin{equation}\label{eqn:available-potential-energy}
\begin{aligned}
    \dv{E_\mathrm{AP}}{t} &= \frac{1}{\bar{N}^2}\int_V b'\partial_t b'\dd{V} \\
    &= \frac{1}{\bar{N}^2}\int_V b'\left(-\bm{u}\cdot\grad b' + \bar{N}^2u_z\right)\dd{V} \\
    &= -\frac{1}{\bar{N}^2}\int_V \div\left(\frac{1}{2}b'^2\bm{u}\right) \dd{V} + \int_V(b' u_z) \dd{V} \\
    &= \int_V(b' u_z) \dd{V}.
\end{aligned}
\end{equation}
Summing these two energy components reveals the exact conservation of total available energy, 
\begin{equation}
    \dv{E_\mathrm{K} + E_\mathrm{AP}}{t} = 0
\end{equation}
for an isolated wave or generalized decaying field.

When analyzing the actual disturbance fields ($\bm{u}'$, $b'$) superimposed on a steady background shear $\bar{\bm{U}} = r\bar{\Omega}(r)\dir{\theta}$, the total disturbance energy is no longer strictly conserved due to the continuous exchange of momentum with the base flow. The evolution of the disturbance kinetic energy evaluates to
\begin{equation}
\begin{aligned}
\dv{E_\mathrm{K}'}{t} &= \int_V\left\{ - u'_z b' -\bm{u}'\cdot\left((\bar{\bm{U}}\cdot\nabla)\bm{u}' + (\bm{u}'\cdot\nabla)\bar{\bm{U}}\right) \right\}\dd{V} \\
&= \int_V\left\{ - u'_z b' -\bm{u}'\cdot\left(-\bar{\Omega}u'_\theta\dir{r}+u'_r \dv{r} (r\bar{\Omega})\dir{\theta}\right) \right\}\dd{V} \\
&= \int_V\left\{ - u'_z b' - \left(r\dv{\bar{\Omega}}{r}\right) u'_r u'_\theta \right\}\dd{V}.
\end{aligned}
\end{equation}
The final term represents the Reynolds stress production, which dictates the transfer of kinetic energy between the background shear and the disturbance. Since the background field is purely two-dimensional, $u_z = u'_z$, and the evolution of the disturbance available potential energy remains the same as (\ref{eqn:available-potential-energy}).

For completeness, we note why the standard potential energy definition, $E_\mathrm{P} = -\int_V z b \dd{V}$, does not form a simple conservation law in this periodic geometry. Multiplying the full buoyancy equation by $z$ yields
\begin{equation}
\begin{aligned}
    \dv{E_\mathrm{P}}{t} &= -\int_V z(\bm{u}\cdot\grad b) \dd{V} \\
    &= -\int_V \div{\left(b z\bm{u}\right)}\dd{V} + \int_V b u_z \dd{V}.
\end{aligned}
\end{equation}
Contrary to the kinetic energy derivation, the divergence term does not vanish upon integration. While the velocity $\bm{u}$ and buoyancy $b$ are strictly periodic in $z$, the full flux vector $b z \bm{u}$ is not, due to the explicit presence of the coordinate $z$. Therefore, as the presence of the non-zero boundary flux prevents the volume integration from cancelling, we have to use available potential energy for periodic domains instead.

Lastly, for an ideal inviscid fluid, the total axial angular momentum of the disturbance, $\mathcal{L}_z$, is a strictly conserved quantity. Integrating the azimuthal momentum equation multiplied by $r$ gives
\begin{equation}
    \dv{\mathcal{L}_z}{t} = -\int_V r ((\bm{u}\cdot\nabla)\bm{u})_\theta \dd{V} - \int_V 2\Omega r u_r \dd{V} - \int_V\frac{1}{\rho_0}\frac{\partial p'}{\partial \theta} \dd{V}.
\end{equation}
Each term on the right-hand side integrates identically to zero. The pressure gradient vanishes because it is the azimuthal derivative of a single-valued function evaluated over a $2\pi$-periodic domain. The advection term vanishes because it can be rewritten as the divergence of the angular momentum flux, $-\div(r u_\theta \bm{u})$, which evaluates to zero over the boundaries. Crucially, the Coriolis term vanishes as a direct mathematical consequence of divergence-free condition. Sepcifically, integrating $\div{\bm{u}} = 0$ over any cylindrical volume of radius $R$ demands that $\int_0^{2\pi}\int_0^{L_z}u_r(R,\theta,z)R\,d\theta \dd{z} = 0$. Because the horizontal average of $u_r$ is identically zero at every radius, the volume integral $\int_V 2\Omega r u_r \dd{V}$ evaluates exactly to zero, thereby guaranteeing the conservation of total axial angular momentum.

\section{Calculation of energy and momentum with spectral decomposition}\label{app:energy}
Having established the theoretical conservation laws for the disturbance fields, we now detail their practical numerical evaluation within the mapped pseudo-spectral framework. To compute the domain-integrated quantities, the physical fields must be evaluated using their discrete spectral representations. 

We begin by considering the available potential energy, which requires integrating the square of the buoyancy perturbation, $b'(r,\theta,z)$. In our spectral framework, this field is represented as a double Fourier series:
\begin{equation}
    b'(r,\theta,z) = \sum_{m=-\infty}^\infty \sum_{\Tilde{k}=-\infty}^\infty b'_{m\Tilde{k}}(r)\exp{\left[\iu \left(m\theta+\frac{2\pi}{L_z}\Tilde{k}z\right)\right]},
\end{equation}
where $m$ and $\Tilde{k}$ are integers representing the azimuthal and axial wavenumbers. The integer $\Tilde{k}$ is utilized explicitly to enforce the periodic boundary condition along the axial direction of length $L_z$. The total available potential energy is defined as the volume integral of $(b')^2$ scaled by the background stratification. Applying orthogonality in the periodic azimuthal and axial directions reduces the three-dimensional integral to a sum of one-dimensional radial integrals:
\begin{equation}
\begin{aligned}
    {E_\mathrm{AP}} &\equiv \frac{1}{2\bar{N}^2} \int_0^{L_z} \int_0^{2\pi} \int_{0}^{\infty} (b')^2 r \dd r \dd\theta \dd z \\
    &= \frac{2\pi L_z}{\bar{N}^2} \sum_{\Tilde{k}}\left\{ \sum_{m > 0}\int_0^\infty |b'_{m\Tilde{k}}(r)|^2 r \dd r + \frac{1}{2}\int_0^\infty |b'_{0\Tilde{k}}(r)|^2 r \dd r \right\},
\end{aligned}
\end{equation}
where we have utilized the conjugate symmetry for real-valued fields, $b'_{-m,-\Tilde{k}}(r) = b'^*_{m\Tilde{k}}(r)$. 

To evaluate the continuous integral over $r$ strictly within the spectral space, we apply the algebraic mapping (\ref{eqn:algebraic-mapping}) of the mapped associated Legendre polynomials, which can be rewritten as: 
\begin{equation}
    r = L\sqrt{(1+\mu)/(1-\mu)},
\end{equation}
where $\mu \in [-1, 1)$ and $L$ is the mapping scale parameter. The differential area element transforms as 
\begin{equation}
    r\dd r = \frac{L^2}{(1-\mu)^2}\dd \mu,
\end{equation}
allowing us to evaluate the radial integrals exactly utilizing Gauss-Legendre quadrature. By sampling at the collocation points $\mu_n$ with corresponding quadrature weights $W_n$, the total available potential energy is computed numerically as
\begin{equation}
    {E_\mathrm{AP}} = \frac{2\pi L_z L^2}{\bar{N}^2} \sum_{\Tilde{k}}\sum_n \left\{ \sum_{m > 0}\frac{|b'_{m\Tilde{k}}(\mu_n)|^2 }{(1-\mu_n)^2} + \frac{1}{2}\frac{|b'_{0\Tilde{k}}(\mu_n)|^2 }{(1-\mu_n)^2} \right\} W_n.
\end{equation}

The above quadrature approach applies identically to the volumetric energy exchange between the disturbance buoyancy and the vertical velocity. Representing the vertical velocity $u_z$ as a double Fourier series analogous to the buoyancy, the integrated total energy exchange evaluates to
\begin{equation}
    \int_V (b' u_z) \dd{V} = 4\pi L_z L^2 \sum_{\Tilde{k}}\sum_n \left\{ \sum_{m > 0}\Re\left[ \frac{b'_{m\Tilde{k}}(\mu_n) u^*_{z,m\Tilde{k}}(\mu_n)}{(1-\mu_n)^2} \right] + \frac{1}{2}\Re\left[ \frac{b'_{0\Tilde{k}}(\mu_n) u^*_{z,0\Tilde{k}}(\mu_n)}{(1-\mu_n)^2} \right] \right\} W_n.
\end{equation}

While the available potential energy and the energy exchange rely on straightforward scalar integrals, evaluating the total kinetic energy requires handling the full perturbation velocity vector. This is achieved efficiently by exploiting the poloidal-toroidal decomposition (\ref{eqn:PT-decomp}). Let
\begin{equation}
    \bm{u}' = \bm{u}_\psi + \bm{u}_\chi.
\end{equation}
The total kinetic energy expands into the sum of the poloidal energy, the toroidal energy, and their cross interaction:
\begin{equation}
    {E_\mathrm{K}} = \frac{1}{2}\int_V (\bm{u}_\psi + \bm{u}_\chi)^2 \dd{V} = \frac{1}{2}\int_V \left( \bm{u}_\psi^2 + \bm{u}_\chi^2 + 2\bm{u}_\psi\cdot\bm{u}_\chi \right) \dd{V}.
\end{equation}
The cross interaction integrates strictly to zero for decaying and periodic fields. Using standard vector identities and the property $\curl\dir{z}=0$, the integrand of the cross interaction expands as $\bm{u}_\chi\cdot\bm{u}_\psi = \div{[(\partial_z\psi)(\grad\chi\cross\dir{z})]}$. Because this interaction is a pure divergence, its volume integral evaluates to zero by Gauss's theorem. The contributions from the strictly poloidal and toroidal fields simplify independently via integration by parts. The toroidal contribution reduces to 
\begin{equation}
    \int_V\bm{u}_\chi^2 \dd{V} = -\int_V \chi \laplacian_\perp \chi \dd{V},
\end{equation}
and the poloidal contribution reduces to 
\begin{equation}
    \int_V\bm{u}_\psi^2 \dd{V} = \int_V \laplacian_\perp\psi \laplacian\psi \dd{V}.
\end{equation}
Combining these, the baseline kinetic energy in terms of the scalar potentials is
\begin{equation}
    {E_\mathrm{K}} = \frac{1}{2}\int_V \left( -\chi \laplacian_\perp\chi + \laplacian_\perp\psi \laplacian\psi \right) \dd{V}.
\end{equation}

As mentioned in \S\ref{sec:PT-decomp} and discussed in \citet{matsushima_1997}, specific global flows require logarithmic components in its poloidal-toroidal decompositions for completeness. In particular, the streamfunction decomposes into a localized component and a structurally imposed logarithmic tail: $\chi \equiv \chi_o + \chi_l P_l(r)$, where $P_l(r) \sim \ln(r)$. Inserting this decomposition into the toroidal energy integral produces a self-energy term for the macroscopic tail, given by $-\chi_l^2 \int_V P_l (r^{-1}\partial_r(r\partial_r P_l)) \dd{V}$. Integrating this term over an unbounded radial domain yields a logarithmic divergence. To obtain a physically meaningful measure of the perturbation energy that characterizes the instability, we evaluate the regularized kinetic energy. This regularization discards the divergent self-energy of the unperturbed $1/r$ tail while explicitly retaining the finite cross interaction that dictates the physical momentum exchange between the localized disturbance $\chi_o$ and the asymptotic boundary structure $\chi_l P_l$:
\begin{equation}
    \widetilde{{E_\mathrm{K}}} = \frac{1}{2}\int_V\left\{ -\chi_o \laplacian_\perp(\chi_o+2\chi_l P_l) + \laplacian_\perp\psi \laplacian\psi \right\} \dd{V}.
\end{equation}

Beyond the energy budgets, the discrete spectral framework also facilitates the exact evaluation of the domain-integrated momenta. The total vertical momentum of the disturbance, $\mathcal{P}_z$, evaluates entirely to the axisymmetric and vertically invariant $m=0, \Tilde{k}=0$ Fourier mode:
\begin{equation}
    \mathcal{P}_z = \int_V u_z \dd{V} = 2\pi L_z\int_0^\infty u_{z,00}(r) r \dd{r} = 2\pi L_z L^2 \sum_n u_{z,00}(\mu_n) \frac{W_n}{(1-\mu_n)^2}.
\end{equation}
Because the vertical velocity is related to the poloidal streamfunction via $u_z = \laplacian_\perp\psi$, we can define a mapped spectral function 
\begin{equation}
    h(\mu) \equiv \laplacian_\perp \psi_{00} / (1-\mu)^2.
\end{equation}
Expanding $h$ in the standard Legendre basis, $h(\mu) = \sum_n h_n P_n(\mu)$, the momentum integral simplifies exactly to the zeroth Legendre coefficient, yielding 
\begin{equation}
    \mathcal{P}_z = 4\pi L_z L^2 h_0.
\end{equation}
Similarly, the total axial angular momentum, $\mathcal{L}_z$, isolates strictly to the zeroth mode of the radial derivative of the toroidal streamfunction:
\begin{equation}
    \mathcal{L}_z = \int_V r u_\theta \dd{V} = \int_V r \left(-\partial_r\chi + \frac{1}{r}\partial_\theta\partial_z\psi\right) \dd{V} = -2\pi L_z\int_0^\infty r^2 \mathrm{d}_r\chi_{00}(r) \dd{r}.
\end{equation}

Finally, the spectral formulation explicitly demonstrates the non-conservative nature of the standard potential energy in a periodic domain, as anticipated theoretically. If evaluated directly, the standard potential energy ($K_\mathrm{P}$) retains an explicit dependence on the domain boundaries:
\begin{equation}
    K_\mathrm{P} = \int_0^{L_z}\int_0^{2\pi}\int_0^\infty z b'(r,\theta,z) r \dd{r} d\theta \dd{z} = 2\pi \sum_{\Tilde{k}} \left( \int_0^\infty b'_{0\Tilde{k}}(r) r \dd{r} \right) \int_0^{L_z} z \exp\left(\iu \frac{2\pi}{L_z}\Tilde{k}z\right) \dd{z}.
\end{equation}
Integration by parts on the axial coordinate reveals the non-zero boundary evaluation:
\begin{equation}
    \int_0^{L_z} z \exp\left(\iu \frac{2\pi}{L_z}\Tilde{k}z\right) \dd{z} = \left[ \frac{L_z}{\iu 2\pi \Tilde{k}} z \exp\left(\iu\frac{2\pi}{L_z}\Tilde{k}z\right) \right]_0^{L_z} - \int_0^{L_z} \frac{L_z}{\iu 2\pi\Tilde{k}} \exp\left(\iu\frac{2\pi}{L_z}\Tilde{k}z\right) \dd{z} = \frac{L_z^2}{\iu 2\pi \Tilde{k}}.
\end{equation}
Because this term is strictly non-zero for periodic modes where $\Tilde{k} \neq 0$, the standard potential energy inherently depends on the artificial vertical domain scale rather than strictly the internal physics. This reinforces the necessity of utilizing the available potential energy when quantifying the physical evolution of the stratified disturbance.

\section{Mathematical identities in spectral basis}\label{app:math-identities}
In this section, we establish the mathematical identities utilized for evaluating the spectral volume integrations. Specifically, we demonstrate that for generalized vector fields satisfying appropriate boundary conditions, the volume integrals of certain rotational operators evaluate exactly to zero. 

First, consider the volume integral of $\dir{z}\cdot\curl{(\bm{A}\cross\curl{\bm{B}})}$ for two arbitrary vector fields $\bm{A}$ and $\bm{B}$. Utilizing standard vector calculus identities, the axial component of this operator can be expressed exactly as the difference of two pure divergence terms:
\begin{equation}
    \dir{z}\cdot\curl{(\bm{A}\cross\curl{\bm{B}})} = \div{\left(A_z\curl{\bm{B}}\right)} - \div{\left((\curl{\bm{B}})_z\bm{A}\right)}.
\end{equation}
Integrating this quantity over the domain volume $V$ and applying Gauss's theorem reduces the evaluation strictly to surface boundary fluxes. As noted by \citet{matsushima_1997}, provided that the vector fields are regular at the origin and decay to zero as $r\rightarrow\infty$, the net boundary flux vanishes completely, yielding:
\begin{equation}\label{eqn:vol-int-z-cross-curl}
    \int_{V} \dir{z}\cdot\curl{(\bm{A}\cross\curl{\bm{B}})} \dd{V} = 0.
\end{equation}

Furthermore, in the specific case where $\bm{A}$ represents a strictly azimuthal background flow ($\bm{A} = A_\theta(r)\dir{\theta}$), the axial component $A_z$ is zero, and the remaining divergence term simplifies to $-r^{-1}\partial_\theta((\curl{\bm{B}})_z A_\theta(r))$. When integrated over the full cylindrical volume, the expression (\ref{eqn:vol-int-z-cross-curl}) also evaluates exactly to zero due to azimuthal periodicity, irrespective of whether $A_\theta(r)$ decays in the far field.

Similarly, we evaluate the domain integral involving the background rotation, shear, and buoyancy. For a solenoidal perturbation velocity $\bm{u}'$ and buoyancy perturbation $b'$, we define the composite vector field 
\begin{equation}
    \bm{V} = \left(2\Omega+f(r)\right)\dir{z}\cross\bm{u}'+b'\dir{z}.
\end{equation}
Utilizing the identity $\dir{z}\cdot\curl{\bm{V}} = \div{(\bm{V}\cross\dir{z})}$ and applying the vector triple product, the axial curl reduces analytically to a strictly horizontal divergence:
\begin{equation}
    \dir{z}\cdot\curl{\bm{V}} = \div{\left[ (2\Omega+f(r))\bm{u}'_\perp \right]},
\end{equation}
where $\bm{u}'_\perp = u'_r\dir{r} + u'_\theta\dir{\theta}$ represents the horizontal velocity. Integrating this exact divergence over the domain volume yields:
\begin{equation}
    \int_V\dir{z}\cdot\curl{\left((2\Omega+f(r))\dir{z}\cross\bm{u}'+b'\dir{z}\right)} \dd{V} = \int_V \div{\left[ \left(2\Omega+f(r)\right)\bm{u}'_\perp \right]} \dd{V}.
\end{equation}
By the divergence theorem, this volume integral reduces entirely to horizontal boundary fluxes evaluated at $r=0$ and $r\rightarrow\infty$. Because the physical velocity field must be regular at the origin and decay to zero in the far field, the radial boundary fluxes are identically zero. Consequently, the integral evaluates to zero independent of the axial boundary conditions.

\section{Linear stability equation of a stratified columnar vortex}\label{app:evp}

The linearized system (\ref{eqn:evp-formulation}) of a stratified azimuthal shear flow can be expressed in terms of the primitive variables:
\begin{equation}
\begin{aligned}
    \dir{r}:
    &&\, \Phi\Tilde{u}_r &= 2(\bar{\Omega}+\Omega)\Tilde{u}_\theta - \mathrm{d}_r \Tilde{p}, \\
    \hat{\bm{\theta}}:
    &&\, \Phi\Tilde{u}_\theta &= -\left(\bar{\zeta}+2\Omega\right)\Tilde{u}_r-\frac{\iu m}{r}\Tilde{p}, \\
    \dir{z}:
    &&\, \Phi\Tilde{u}_z &= -\Tilde{b} - \iu k \Tilde{p}, \\
    \Tilde{b}:
    &&\, \Phi\Tilde{b} &= \bar{N}^2\Tilde{u}_z,
\end{aligned}\label{eqn:matrix-evp}
\end{equation}
where $\bar{\Omega}$ and $\bar{\zeta}$ are the angular velocity and axial vorticity of the background shear, and $\Phi(r)\equiv \sigma + \iu m\bar{\Omega}$ is the Doppler-shifted complex eigenvalue. 
This system of equations can be simplified to a second-order eigenvalue problem of $g(r)\equiv r\Tilde{u}_r$:
\begin{equation}
\begin{aligned}
    0 =&& g'' &+ \left\{\frac{2\Phi\Phi'}{\Phi^2+\bar{N}^2}+\frac{\iu m (2\bar{\Omega}-\bar{\zeta})}{r\Phi}+\frac{\Phi'}{\Phi}+\frac{1}{r}-\frac{S'}{S}\right\}g' \\
    && &- \left\{\frac{k^2 \cdot(\Phi^2+\Delta(r))}{\Phi^2+\bar{N}^2}+\frac{m^2}{r^2}+\frac{2\iu m(\bar{\zeta}+2\Omega)\Phi'}{r(\Phi^2+\bar{N}^2)}+\frac{\iu m[S\bar{\zeta}'-(\bar{\zeta}+2\Omega)S']}{r\Phi S}\right\}g,
\end{aligned}
\label{eqn:2nd-order-linear}
\end{equation}
where $S(r)\equiv (k^2r^2+m^2)\Phi^2+m^2\bar{N}^2 (\geq 0)$.



A quick inspection of equation~(\ref{eqn:2nd-order-linear}) shows two types of critical-layer singularities on top of the pole ($r = 0$) singularity:
\begin{enumerate}
    \item The \textit{Baroclinic critical layer}, where 
        \begin{equation}
            \Phi^2(r_c) + \bar{N}^2 \equiv 0,
        \end{equation}
        so $\sigma_c = -\iu m \bar{\Omega}(r_c) \pm \iu\bar{N}$;
    \item The \textit{Barotropic critical layer}, which corresponds to the case
        \begin{equation}
            \Phi(r_c) \equiv 0,
        \end{equation}
        so $\sigma_c = - \iu m \bar{\Omega}(r_c)$, which gives a continuous spectrum of the eigenvalues.
\end{enumerate}
\printcredits

\bibliographystyle{cas-model2-names}

\bibliography{reference}



\end{document}